\documentclass[twocolumn,floatfix]{revtex4}%
\usepackage[dvipdfmx]{graphicx}%
\usepackage{amsmath}%
\usepackage{amsfonts}
\usepackage{amssymb}
\usepackage{hhline}
\usepackage{bm}
\usepackage{mathrsfs}
\usepackage{color}

\def\e{{\epsilon}}
\def\k{{ {\bm k} }}

\def\q{{ {\bm q} }}

\def\0{{ {\bm 0} }}
\def\w{{\omega}}
\def\a{{\alpha}}

\allowdisplaybreaks[4]

\begin{document}
\title{Fully gapped $s$-wave superconductivity\\ 
enhanced by magnetic criticality in heavy fermion system
}
\author{
Rina Tazai,  and
Hiroshi Kontani
}

\date{\today }

\begin{abstract}
In heavy fermion systems,
higher-rank multipole operators are active thanks to the
strong spin-orbit interaction (SOI),
and the role of diverse multipole fluctuations
on the pairing mechanism attracts a lot of attention.
Here, we study a mechanism of superconductivity 
in heavy fermion systems, by focusing on the
impact of vertex corrections (VCs) for the pairing interaction
going beyond the Migdal approximation.
In heavy fermion systems, strong interference between
multipole fluctuations cause significant VCs, that represent
many-body effects beyond mean-field-type approximations.
Especially, the coupling constants between electrons and charged-bosons,
including the electron-phonon coupling constant,
are strongly magnified by the VCs.
For this reason,
moderate even-rank (=electric) multipole fluctuations 
give large attractive interaction,
and therefore $s$-wave superconductivity 
can emerge in heavy-fermion systems.
In particular, phonon-mediated superconductivity
is expected to be realized near the magnetic criticality,
thanks to the VCs due to magnetic multipole fluctuations.
The present mechanism may be responsible for the 
fully gapped $s$-wave superconducting state
realized in CeCu$_2$Si$_2$.

\end{abstract}

\address{
Department of Physics, Nagoya University,
Furo-cho, Nagoya 464-8602, Japan. 
}
 

\sloppy

\maketitle

\section{Introduction}
\label{sec:Intro}
Heavy fermion systems are very interesting 
platform of exotic electronic states induced by
strong Coulomb interaction and spin-orbit interaction (SOI)
on $f$-electrons.
In Ce-based compounds, $4f^1$ configuration is 
realized in Ce$^{3+}$ ion.
Due to strong SOI, the total angular momentum $J=L+S$
becomes good quantum number.
Since the energy of $J=5/2$ multiplet is about 0.3eV lower than 
that of $J=7/2$ multiplet,
the latter can be safely dropped in the theoretical model.
In tetragonal crystals, the degeneracy of $J=5/2$ multiplet is separated
into three Kramers doubles 
due to the crystalline electric field (CEF).
Usually, the CEF splitting energy is of order
1 meV-10 meV.

In many $f$-electron systems, magnetic fluctuations cause interesting 
quantum critical phenomena and 
unconventional superconductivity
\cite{Moriya,Yamada,Kontani-rev,Monthoux,Dahm,Scalapino,Takimoto-SC}.
In addition, higher-rank multipole operators are also active thanks to the
strong SOI of $f$-electrons.
For this reason, various interesting multipole order and fluctuations
are caused by strong $f$-electron interaction.
As an example of higher-rank order, CeB$_6$ exhibits quadrupole (rank 2) 
order and field-induced octupole (rank 3) order \cite{Shiina,Kusunose}.
Also, emergence of hexadecapole (rank 4) in PrRu$_4$P$_{12}$ \cite{Takimoto}
and hexadecapole or dotriacontapole (rank 5) in URu$_2$Si$_2$ 
\cite{Haule,Oka,Ikeda-Uru2}
have been discussed.
The fluctuations of these multipole operators
mediate interesting unconventional superconductivity.
For example,
$d$-wave superconductivity appears next to the 
the magnetic order phase in Ce$M$In$_5$ ($M$=Rh,Co,Ir)
\cite{Izawa-115}.
In addition, superconductivity appears next to the quadrupole order
in Pr$T_2$Zn$_{20}$ ($T$ = Rh and Ir)  \cite{Oni-Pr}
and Pr$T_2$Al$_{20}$ ($T$=V,Ti) \cite{Naka-Pr}.
These Pr-based superconductors indicate that 
the higher-rank ($\ge 2$) multipole fluctuations inherent in $f$-electron systems
mediate exotic superconducting states $f$-electron systems.

CeCu$_2$Si$_2$ is the first discovered heavy-fermion 
superconductor
\cite{Ste-122,Yuan-122},
and its discovery triggered huge amount of research on 
unconventional superconductivity in various compounds
\cite{Peid}.
At ambient pressure,
CeCu$_2$Si$_2$ shows superconducting transition at $T_{\rm c}\approx0.6$K
near the magnetic instability
\cite{Ishida}.
Under pressure, $T_{\rm c}$ suddenly increases to $1.5$K at $P_c\approx4.5$GPa.
For long time, 
CeCu$_2$Si$_2$ has been considered as a typical $d$-wave superconductor
mediated by magnetic fluctuations.
However, $d$-wave nodal gap structure contradicts with 
exponentially small specific heat at $T\ll T_{\rm c}$ as reported in Refs.\cite{Kit1-122,Kit2-122}.
Later, the fully gapped state
is confirmed by the measurements of thermal conductivity 
and penetration depth at very low temperatures \cite{Yama-122,Steglich}.
In addition, the robustness of $T_{\rm c}$ against randomness 
indicates that plain $s$-wave superconductivity 
without sign-reversal is realized in CeCu$_2$Si$_2$
\cite{Yama-122}.

It is a significant challenge for theorists 
to establish a realistic microscopic theory of 
fully gapped $s$-wave superconductivity in heavy-fermion systems,
against large Coulomb repulsion.
It is believed that fluctuations of even-rank multipole operators, 
such as charge, quadrupole and hexadecapole operators,
mediate attractive pairing interaction.
To realize large even-rank multipole fluctuations,
at least two Kramers doublets should contribute to the Fermi surface,
if the charge (rank 0) fluctuations are 
suppressed by Coulomb interaction.
In fact, in CeCu$_2$Si$_2$ at ambient pressure,
two Kramers doublets form the Kondo resonance below 10K
according to the first-principles study based on the LDA+DMFT 
\cite{LDADMFT_multiporbital}.
Pressure-induced change in multiorbital nature
may be a key to understand the 
$P$-$T$ phase diagram in  CeCu$_2$Si$_2$
\cite{LDADMFT_multiporbital,Hat-122,Hol-122}.


In the random-phase-approximation (RPA),
even-rank multipole fluctuations are always smaller than odd-rank ones.
Therefore the obtained gap structure inevitably 
possesses sign reversal within the Migdal approximation
\cite{Ikeda-122}.
This discrepancy indicates the significance of 
higher-order many-body effects
called the vertex corrections (VCs).
In fact, the VC for the electron-boson coupling,
which we call $U$-VC, has been studied in Refs.
\cite{Schrieffer,Stamp,Dahm,Yamada,Kontani-rev,Onari-SCVC,Onari-FeSe,Yamakawa-FeSe}.
The violation of Migdal theorem \cite{Migdal} due to the Maki-Thompson (MT) and
Aslamazov-Larkin (AL) VCs, which are respectively the first-order
and second-order corrections with respect to the susceptibility,
have been studied in Refs. \cite{Stamp,Dahm,Onari-SCVC,rina2}.
In multiorbital systems,
moderate orbital fluctuations induce
strong attractive pairing interaction thanks to the 
AL-type $U$-VC
\cite{rina2,Yamakawa-FeSe2}.
However, strong SOI in $f$-electron systems
has prevented the detailed analysis of the VCs.
Thus, it is highly required to 
construct the theoretical formalism to 
analyze the VCs in systems with strong SOI.
We stress that the DMFT has been successfully 
applied to $f$-electron systems 
\cite{Haule,LDADMFT_multiporbital,KotGeo-DMFT,KotVol-DMFT,Held,DMFT-HF,Otsuki-HF},
while strong $k$-dependence of VCs near the 
magnetic quantum-critical-point (QCP)
is not fully taken into consideration.

In this paper, we propose a mechanism of $s$-wave superconductivity
in multi-orbital heavy fermion systems by focusing on the VCs
beyond Migdal approximation.
%
Near the magnetic QCP,
various types of multipole fluctuations
develop simultaneously,
due to the combination of strong SOI and Coulomb interaction.
The developed multipole fluctuations give significant VCs
in heavy fermion systems.
Especially, the VCs significantly magnify the 
attractive pairing interaction due to even-rank multipole fluctuations,
so the Migdal theorem is no more valid.
Due to this mechanism, $s$-wave superconductivity can be realized 
in heavy fermion systems,
once moderate (phonon-induced) quadrupole or hexadecapole fluctuations exist.
The $s$-wave superconductivity is strongly enhanced near the magnetic criticality.
The present mechanism may be responsible for the 
fully gapped superconducting state
realized in CeCu$_2$Si$_2$.

In 3$d$-electron systems, the AL-type VCs are
efficiently calculated by using the SU(2) symmetry
in the spin-space.
Thus, the same formalism cannot be applied to 5$d$ or 
$f$-electron systems because of the violation of SU(2) symmetry.
To overcome this difficulty, 
we introduce a natural two-orbital periodic Anderson model, in which the pseudo-spin 
of $f$-electron satisfies the axial rotational symmetry.
By virtue of this fact,
we can analyze complicated VCs efficiently.
In the present model, 16 type multiple operators 
(rank 0$\sim$5) are active, so we can discuss  
rich physics associated with higher-rank multipole operators.


\section{Model}
\label{sec:Model}

In this section, we derive an useful two-orbital periodic Anderson model (PAM)
for CeCu$_2$Si$_2$, in which we can define the pseudo-spin
that satisfy the conservation law.
For this purpose, we first introduce a general three-orbital 
$J=5/2$ PAM for describing $4f^{1}$ electrons in $\rm{Ce}$-based compounds.
The kinetic term is given by
\begin{eqnarray}
\hat{H}_{0}^{\rm{general}}&=&\sum_{\k\sigma}\epsilon_{\k}c^{\dagger}_{\k\sigma}c_{\k\sigma}+\sum_{\k l\Sigma}E_{l}f^{\dagger}_{\k l\Sigma}f_{\k l\Sigma} \nonumber \\
&+&\sum_{\k l\sigma\Sigma}\left(V^{*}_{\k l\sigma\Sigma}f^{\dagger}_{\k l\Sigma}c_{\k \sigma}
+V_{\k l\sigma\Sigma}c^{\dagger}_{\k\sigma}f_{\k l\Sigma}\right)
\label{eqn:hamiltonian}
\end{eqnarray}
where $c^{\dagger}_{\k \sigma}$ ($c_{\k \sigma}$) is a creation 
(annihilation) operator for $s$-electron with momentum $\k$, spin $\sigma$
and energy $\epsilon_{\k}$.
$f^{\dagger}_{\k l \Sigma}$ ($f_{\k l \Sigma}$) is a creation (annihilation) operator for 
$f$-electron with $\k$, orbital $l$ ($l=1,2,3$), pseudo-spin $\Sigma$,
and energy $E_{l}$.
$V_{\k l\sigma\Sigma}$ is the 
hybridization term between $f$ and $s$ electrons.

\begin{figure}[htb]
\includegraphics[width=.96\linewidth]{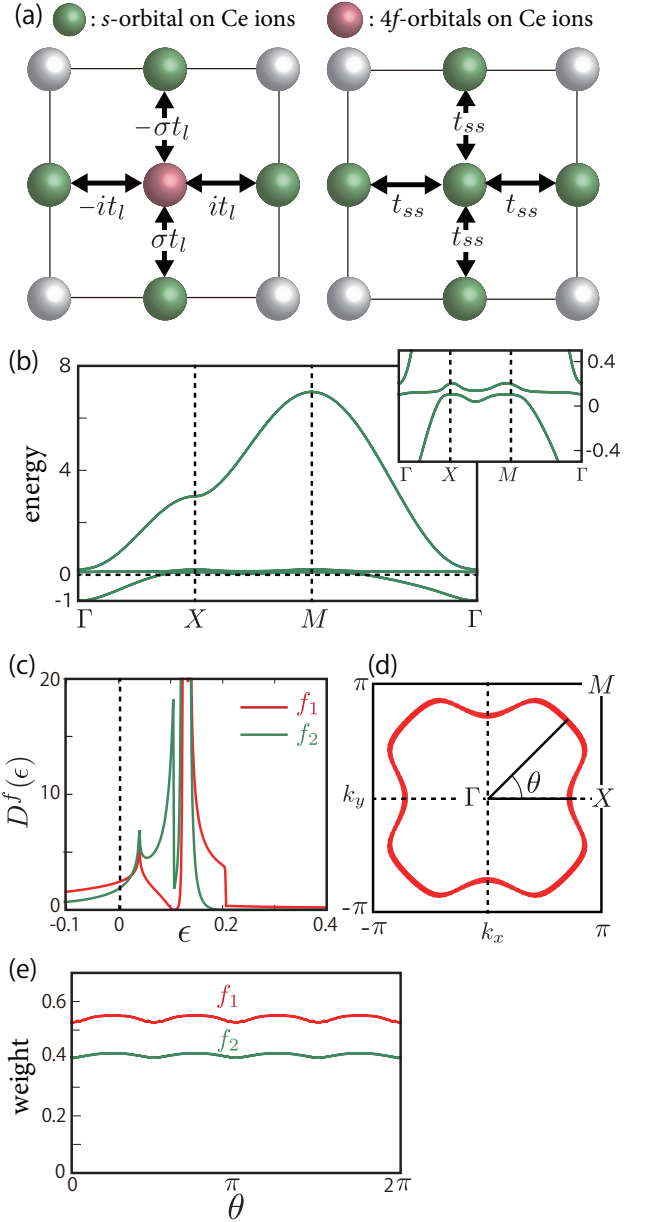}
\caption{(color online)
(a) The nearest neighbor hopping integrals given by $s$-$s$ and $s$-$f$ hopping.
$\sigma=1(-1)$ for pseudo-spin up (down), $t_{1}=-\sqrt{3/14}\,\,t_{sf}$, $t_{2}=-t_{1}/\sqrt{5}$.
(b) Band dispersion along high-symmetry line.
(c) Partial DoS of $f_{l}$-electrons. 
The red (green) line corresponds to $f_{1}$($f_{2}$)-orbital.
(d) Obtained Fermi surface. (e) $\theta$-dependence of the 
$f_{l}$-orbital weight on Fermi surface. 
The red (green) line corresponds to $f_{1}$($f_{2}$)-orbital.}
\label{fig:band}
\end{figure}

Here, we derive an useful simplified PAM for CeCu$_2$Si$_2$ from Eq. (\ref{eqn:hamiltonian}). 
According to the LDA+DMFT study for $\rm{CeCu_{2}Si_{2}}$\cite{LDADMFT_multiporbital},
the following two Kramers doublets give dominant DoS around the Fermi energy at
ambient pressure.
They are expressed  in the $J_{z}$ basis as,
\begin{eqnarray}\,&&
\begin{cases}
|f_{1}\Uparrow \rangle&=a|- \frac{5}{2}\rangle+b|+\frac{3}{2}\rangle,\nonumber \\
|f_{1}\Downarrow \rangle &=a|+ \frac{5}{2}\rangle+b|-\frac{3}{2}\rangle , 
\end{cases}  \\  \,&& 
\begin{cases} |f_{2}\Uparrow \rangle&=-a|+ \frac{3}{2}\rangle+b|- \frac{5}{2}\rangle, \\
|f_{2}\Downarrow \rangle&=-a|- \frac{3}{2}\rangle+b|+ \frac{5}{2}\rangle, \end{cases} 
\label{eqn:wavefunc}
\end{eqnarray}
where $\Uparrow(\Downarrow)$ denotes pseudo-spin up (down) of $f_{l}$-electron ($l=1,2$).
We drop the third Kramers doublet $|f_{3} \rangle =|J_{z}=\pm \frac{1}{2} \rangle$ 
, since it gives negligibly small weight near the Fermi level.
We study 2D square lattice model as shown in Fig.\ref{fig:band}(a). 
Both $f$- and $s$-orbital are on Ce ion.
For simplicity, we consider only the above-mentioned two-orbitals.
We introduce only the nearest neighbor
$s$-$f$ and $s$-$s$ hopping integrals.
In this case, $f$-electron with pseudo-spin $\Uparrow(\Downarrow)$
hybridizes with only $s$-electron with $\uparrow (\downarrow)$ as we confirm in Appendix A.
Thus, the pseudo-spin is conserved, and we can put $\Sigma=\sigma$.
In the present two-orbital model, the kinetic term is given by
\begin{eqnarray}
\hat{H}_{0}&=&\sum_{\k\sigma}\epsilon_{\k}c^{\dagger}_{\k\sigma}c_{\k\sigma}+\sum_{\k l\sigma}E_{l}f^{\dagger}_{\k l\sigma}f_{\k l\sigma}\nonumber \\
&+&\sum_{\k l\sigma}
\left(V^{*}_{\k l\sigma}f^{\dagger}_{\k l\sigma}c_{\k\sigma}+V_{\k l\sigma}c^{\dagger}_{\k\sigma}f_{\k l\sigma}\right) \nonumber \\
&=&\sum_{\k\sigma}\hat{a}^{\dagger}_{\k\sigma}\hat{h}_{\k}^{\sigma} \hat{a}_{\k\sigma},
\label{eqn:hamiltonian2}
\end{eqnarray}
where $\sigma$ is the real (pseudo) spin for $s$- ($f$-)electron and
 $\hat{a}^{\dagger}_{\k\sigma}\equiv(f^{\dagger}_{\k1\sigma},f^{\dagger}_{\k2\sigma},c^{\dagger}_{\k\sigma})$.
By using the Slater-Koster tight-binding method \cite{Takegahara,Saso}, the $s$-$f$ hybridizations are given as
\begin{eqnarray}
V_{\k f_{1}\uparrow}&=&-\sqrt{\frac{3}{14}}t_{sf}(a\sqrt{5}+b)(\sin k_{y} - i\sin k_{x} ),\nonumber \\
V_{\k f_{1}\downarrow}&=&\sqrt{\frac{3}{14}}t_{sf}(a\sqrt{5}+b)(\sin k_{y} + i\sin k_{x} ),\nonumber \\
V_{\k f_{2}\uparrow }&=&\sqrt{\frac{3}{14}}t_{sf}(a-\sqrt{5}b)(\sin k_{y} - i\sin k_{x} ),\nonumber \\
V_{\k f_{2}\downarrow}&=&-\sqrt{\frac{3}{14}}t_{sf}(a-\sqrt{5}b)(\sin k_{y} + i\sin k_{x} ).
\label{eqn:hopping}
\end{eqnarray}
Hereafter, we simply put  $a=1$,
$b (=\!\!\sqrt{1-a^2})\!\!=\!0$. Actually, the relation $a\simeq 1$ is 
reported by recent resonant X-ray scattering experiment in Ref.\cite{RIXS}.
In this case, we obtain $|V_{\k f_{1}\sigma}/V_{\k f_{2}\sigma}|=\sqrt{5}$.
Thus, $f_{1}$-orbital is more itinerant than $f_{2}$-orbital.
This feature is consistent with the results of previous DMFT calculation 
for CeCu$_2$Si$_2$ in Ref.\cite{LDADMFT_multiporbital}, which
show $V_{\k f_{2}\sigma}\approx 2V_{\k f_{1}\sigma}$.
The schematic picture of the $s$-$s$ and $s$-$f$ hopping integrals are 
shown in Fig.\ref{fig:band}(a).
We fix the parameters $\epsilon_{\k}=2t_{ss}(\cos k_{x}+\cos k_{y})+\epsilon_{0}$, $t_{ss}=-1.0$, 
$\epsilon_{0}=3.0$, $t_{sf}=0.7$, and $f$-electron energy $E_{f_{1}}=0.2$ and $E_{f_{2}}=0.1$. 
We set the temperature $T=0.02$ and the chemical potential 
$\mu=-5.52\times10^{-3}$ in the following numerical study.
Then, $f$-electron number is $n_{f}=0.9$,
and $s$-electron number is $n_{s}=0.3$. 

In Fig.\ref{fig:band}(b), we show the obtained band structure.
$\epsilon=0$ corresponds to the chemical potential.
In the present 3 band model, the lowest band crosses the Fermi level.
The total band width is $W_{D}\sim 10$ (in unit $|t_{ss}|=1$).
$|t_{ss}|$ is of order 1eV since $W_{D}\sim 10$eV in CeCu$_2$Si$_2$ \cite{Ikeda-122}.
The width of quasi-particle band (=the lowest band) is 
$W_{D}^{qp} \sim 1$.
Density of states (DoS) for $f_{l}$-orbital;
$D^{f_{l}}(\epsilon)$ is given in Fig.\ref{fig:band}(c).
Here, the relation $D^{f_{1}}(0)\simeq D^{f_{2}}(0)$ is satisfied.
In the present study, we neglect self energy.
Figure \ref{fig:band}(d) shows the obtained Fermi surface.
In Fig.\ref{fig:band}(e), we plot the $\theta$-dependence of the $f_{l}$-orbital weight, 
where $\theta$ is angle of the Fermi momentum defined in Fig.\ref{fig:band}(d).
We stress that the weights of $f_{1}$- and $f_{2}$-orbital are comparable regardless of $\theta$, which originates from the isotropic $s$-$f$ hybridization given in Eq. (\ref{eqn:hopping}) due to the strong SOI.
(In contrast, in 3$d$-electron system such as Fe-based compounds, 
the $d$-orbital weight shows strong $\theta$-dependence.)
This fact is favorable for the development of multiple higher-rank multipole susceptibilities, as we will show in Sec. IV.

If we consider the $f$-$f$ hopping,
the $f_{l}$-orbital weight comes to have $\theta$-dependence.
Even in this case, the multiple higher-rank multipole susceptibilities can develop when $t_{ff} \ll t_{sf}$, which is naturally expected  in heavy fermion compounds.
We will discuss this in more detail in Appendix D and in the future publication \cite{Tazai-future}.

We introduce on-site Coulomb interaction in $f$-electrons,
\begin{eqnarray}
\hat{H}_{U}=u\cdot\frac{1}{4}\sum_{i,l l' m m'}\sum_{\sigma \sigma' \rho \rho'}
U_{ll';mm'}^{0,\sigma \sigma';\rho\rho'}
f^{\dagger}_{i l\sigma}
f_{i l'\sigma'}
f_{i m\rho} 
f^{\dagger}_{i m'\rho'},
\label{eqn:U0}
\end{eqnarray}
where $i$ is site index, and $u$ is the Coulomb interaction.
$\hat{U}^{0}$ is the interaction matrix normalized on the condition that $U^{0,\sigma \bar{\sigma}; \sigma \bar{\sigma}}_{11;11}\equiv U^{1}=1$.
Note that $\hat{U}^{0}$ in Eq. (\ref{eqn:U0}) is antisymmetrized.

Here, we derive $\hat{U}^{0}$ in Eq. (\ref{eqn:U0}) from 
the following $L_{z}$-basis Coulomb interaction:
\begin{eqnarray}
\bar{U}_{l_{z},l_{z}',l_{z}'',l_{z}'''}^{0}&=&\frac{e^2}{4\pi\epsilon_{0}}\int d\vec{r} d\vec{r'} \frac{u_{l_{z}}^{*}
(\vec{r})u_{l_{z}'''}^{*}(\vec{r'})u_{l_{z}'}(\vec{r'})u_{l_{z}''}(\vec{r})}{{|\vec{r}-\vec{r'}|}} \nonumber \\ 
&=&\sum_{p}a^{p}_{l_{z},l_{z}',l_{z}'',l_{z}'''} F^{p}, \label{eqn:lzcoulomb}
\end{eqnarray} where $u_{l_{z}}(\vec{r})(=\hspace{-5pt}R(r)\Theta_{l_{z}}(\theta)e^{il_{z}\phi})$
is the wave function of the $f$ electron with $l_{z}$ in the absence of 
the SOI.
$F^{p}$ is the Slater integral
introduced in Ref.\cite{Slater U}, which is defined as 
$F^{p}\!\!=\!\! \frac{e^2}{4\pi \epsilon_{0}} \int dr \int dr' R^{2}(r)R^{2}(r') r_{\rm min}^{p} r_{\rm max}^{-(p+1)} r^{2}r'^{2}$, where $r_{\rm min}=\min \{ r,r' \}$ and $r_{\rm max}=\max \{ r,r' \}$.
In this paper, we put $(F^{0},F^{2},F^{4},F^{6})=(5.3,9.09,6.927,4.756)$ in unit eV by referring Ref.\cite{F0F2F4F6}. Finally, we determine $\hat{U}^{0}$ in Eq. (\ref{eqn:U0}) by performing the unitary transformation of Eq. (\ref{eqn:lzcoulomb}) and normalizing it on condition that  $U^{1}=1$.

The present Coulomb interaction in Eq. (\ref{eqn:U0}) does not satisfy SU(2) symmetry in the pseudo-spin space. Nonetheless, the pseudo-spin is conserved in Eq. (\ref{eqn:U0}) for any value of $a$ in Eq. (\ref{eqn:wavefunc}).
Equivalently, $\hat{U}^{0}$ satisfies the axial rotational symmetry along $z$-axis.
Then, $\hat{U}^{0}$ is uniquely decomposed into in-plane spin ($=\!\!s$), out-of-plane spin ($=\!\!s\!\!\perp$) and charge ($=\!\!c$) channels as follows:
\begin{eqnarray}
U_{ll';mm'}^{0;\sigma \sigma' ; \lambda \lambda'}
=\frac{1}{2}U^{0;s}_{ll';mm'}
(\sigma^{x}_{\sigma \sigma'} \sigma^{x}_{\lambda' \lambda}+
\sigma^{y}_{\sigma \sigma'} \sigma^{y}_{\lambda' \lambda})\nonumber \\
+\frac{1}{2}U^{0;s\perp}_{ll';mm'}
\sigma^{z}_{\sigma \sigma'} \sigma^{z}_{\lambda' \lambda}
+\frac{1}{2}U^{0;c}_{ll';mm'}\sigma^{0}_{\sigma \sigma'}\sigma^{0}_{\lambda' \lambda},
 \label{eqn:U0sc}
\end{eqnarray}
where $\vec{\sigma}=(\sigma^{x},\sigma^{y},\sigma^{z})$ is Pauli matrix vector in the pseudo-spin space, and $\sigma^{0}$
is identity matrix.
$\hat{U}^{0;ch} (ch=s,s\!\!\perp, c)$ is defined as
\begin{eqnarray} \begin{cases}
\hat{U}^{0;s}=\hat{U}^{0;\uparrow \uparrow ; \uparrow \uparrow} - \hat{U}^{0; \uparrow \uparrow ; \downarrow \downarrow} \\
\hat{U}^{0;s\perp}=\hat{U}^{0;\uparrow\downarrow ; \uparrow\downarrow}  \\
\hat{U}^{0;c}=\hat{U}^{0;\uparrow\uparrow ; \uparrow\uparrow} + \hat{U}^{0; \uparrow\uparrow ; \downarrow\downarrow}. 
\end{cases}
\end{eqnarray}
The matrix elements of $\hat{U}^{0;ch} (ch=s,s\!\!\perp, c)$ are summarized in TABLE \ref{tab:coulomb}.
Each elements are composed of the intra-orbital Coulomb interaction $U$,
inter-orbital one $U'$, exchange interactions $J, J^{\perp}, J'$, $J^{x1}$, and  $J^{x2}$.
The definition and numerical value of each component are given in Appendix B.
In the case of $a=1$ and $b=0$, the other elements not listed 
in the TABLE \ref{tab:coulomb} become zero.
Although some of these elements (e.g., $U^{0;ch}_{11;12}$) come to be finite for $a \lesssim 1$,
they remain very small and negligible.
 Therefore, TABLE \ref{tab:coulomb} is still useful, practically.
Note that $a=\sqrt{5/6}$ and $b=\sqrt{1/6}$ are satisfied in cubic symmetry.
\begin{table}[tbh]
\begin{minipage}{0.45\hsize}
\begin{center}
\begin{tabular}{|c|c|c|c} \hline
$s$ & type & value  \\ \hline
$U^{0;s}_{11;11}$ & $U^{1}$ & 1.0  \\ \hline
$U^{0;s}_{22;22}$ & $U^{2}$ & $0.90$  \\  \hline
$U^{0;s}_{lm;lm}$ & $U'-J+J^{\perp}$  & 0.80 \\  \hline
$U^{0;s}_{ll;mm}$ & $J-J^{x1}$  & $-$0.12 \\  \hline
$U^{0;s}_{lm;ml}$ & $J'-J^{x2}$  & 0.20 \\  \hline
    \end{tabular}
       \end{center}
 \end{minipage}
\begin{minipage}{0.45\hsize}
\begin{center}
  \begin{tabular}{|c|c|c|c} \hline
$s\!\!\perp$ & type & value \\ \hline
$U^{0;s\perp}_{11;11}$ & $U^{1}$ & 1.0 \\ \hline
$U^{0;s\perp}_{22;22}$ & $U^{2}$ & 0.90\\  \hline
$U^{0;s\perp}_{lm;lm}$ & $U'-J^{x1}$  & 0.68 \\  \hline
$U^{0;s\perp}_{ll;mm}$ & $J^{\perp}$  & 0.0 \\  \hline
$U^{0;s\perp}_{lm;ml}$ & $J'-J^{x2}$  & 0.20 \\  \hline
    \end{tabular}
      \end{center}
\end{minipage}
\begin{center}
  \begin{tabular}{|c|c|c|c} \hline
$c$  & type & value \\ \hline
$U^{0;c}_{11;11}$ & $-U^{1}$ & $-$1.0 \\ \hline
$U^{0;c}_{22;22}$ & $-U^{2}$ & $-$0.90 \\  \hline
$U^{0;c}_{lm;lm}$ & $U'-J-J^{\perp}$  & 0.80 \\  \hline
$U^{0;c}_{ll;mm}$ & $J-2U'+J^{x1}$  & $-$1.5 \\  \hline
$U^{0;c}_{lm;ml}$ & $-J'+J^{x2}$  & $-$0.20 \\  \hline
\end{tabular}
\end{center}
\caption{Matrix elements of Coulomb interaction 
for in-plane spin channel (top left), out-of-plane spin channel (top right),
and charge channel (bottom) for $l\neq m$.
$J=J'$, $J^{\perp}=0$ and $J^{x1}=-J^{x2}$ are satisfied in the present two-orbital model.}
\label{tab:coulomb}
\end{table}

In the present two-orbital model in Eq. (\ref{eqn:wavefunc}), there are 
16-type active multipole operators up to rank 5;
monopole (rank 0 ), dipole (rank 1), quadrupole (rank 2), octupole (rank 3), hexadecapole (rank 4) and dotriacontapole (rank 5)
moment as shown in TABLE {\ref{tab:multipole} \cite{Ikeda-Uru2}.
Some operators belong to the same irreducible representation (IR).
Since the system is inversion symmetric,
an even-rank (odd-rank) operator corresponds to an electric (magnetic) multipole operator.
Each multipole operator of rank $k$ are composed of  $4\times 4$ tensor $J^{(k)}_{q} (q=-k\sim k)$
\cite{Shiina,Springer} which is given by 
 \begin{eqnarray}
 [J_{\pm},J^{(k)}_{q}]&=&\sqrt{(k\mp q)(k\pm q+1)}J^{(k)}_{q\pm1} \\
J_{k}^{(k)}&=&(-1)^{k}\sqrt{\frac{(2k-1)!!}{(2k)!!}}J_{+}^{k}.
 \label{eqn:J3}
\end{eqnarray}
By using $J^{(k)}_{q}$, we obtain $4\times4$ multipole operators $\hat{O}^{Q}$.
Here, $Q\equiv(\Gamma, \phi)$, where $\Gamma$ is index of the irreducible representation
 $(\Gamma=A^{+}_{1},A^{+}_{2},E^{+},A^{-}_{1},A^{-}_{2},E^{-})$ and $\phi$ 
 is index of independent multipole operator $(\phi=1\sim  N_{\Gamma})$. 
 For each $\Gamma$, $N_{\Gamma}$ is given in TABLE \ref{tab:multipole}.
 The matrix representations for 16-type operators
are given in Appendix C.
\begin{table}[htb]
  \begin{tabular}{|c|c|c|c|c|} \hline
    \hspace{3mm}IR ($\Gamma$) \hspace{3mm} & rank (k) & Operator (Q) & $N_{\Gamma}$& $ch_{\Gamma}$ \\ \hhline{|=|=|=|=|=|}
     & $0$  & $\hat{1}$ & &\\ \cline{2-3}
     $A_{1}^{+}$& $2$  & $\hat{O}_{20}$ &3  & $c$ \\ \cline{2-3}
      & $4$  & $\hat{H}_{0}$ & & \\ \hline
    $A_{2}^{+}$ & $4$  & $\hat{H}_{z}$ &1 & $s$ \\   \hline
    $E^{+}$ & $2$  & $\hat{O}_{yz},\hat{O}_{zx}$  &2 & $s_{\perp}$ \\  \hhline{|=|=|=|=|=|}
    $A_{1}^{-}$ & $5$  & $\hat{D}_{4}$ &1 & $c$ \\   \hline
     & $1$  & $\hat{J}_{z}$ & & \\  \cline{2-3}
    $A_{2}^{-}$  & $3$  & $\hat{T}_{z}$ & 3 & $s$ \\  \cline{2-3}
      & $5$  & $\hat{D}_{z}$ & & \\ \hline
       & $1$  & $\hat{J}_{x}$,$\hat{J}_{y}$ & & \\  \cline{2-3}
      $E^{-}$ & $3$  & $\hat{T}_{x}$,$\hat{T}_{y}$ & 6 & $s_{\perp}$ \\  \cline{2-3}
      & $5$  & $\hat{D}_{x}$,$\hat{D}_{y}$  & & \\ \hline
    \end{tabular}
    \caption{Irreducible representation and 16-type active multipole operators in the present two-orbital 
    model. Operator with rank $k$ corresponds to 
$2^{k}$-pole. $N_{\Gamma}$ is the number of operators in symmetry $\Gamma$. 
Each operator is classified into the pseudo-spin or charge channel, $ch_{\Gamma}$.}
    \label{tab:multipole}
\end{table}

Here, we introduce the effective on-site electric multipole-multipole  
interaction $\hat{V}^{\rm ph}$ that belongs to $A_{1}^{+}$ symmetry (= identical representation),
\begin{eqnarray}
V_{ll'mm'}^{\rm ph}&=&2gW_{ll'mm'}\nonumber \\
&=&2g (\hat{C}^{A^{+}_{1}})_{ll'}(\hat{C}^{A^{+}_{1}})_{mm'},
 \label{eqn:ph}
\end{eqnarray}
where $\hat{C}^{A^{+}_{1}}$ is the dimensionless matrix given by a linear combination of multipole operators belong to
$\Gamma=A^{+}_{1}$  in TABLE {\ref{tab:multipole}. It is expressed as
\begin{eqnarray}\hat{C}^{A^{+}_{1}}\equiv 
\alpha \hat{\tau}^{0} +\beta \hat{\tau}^{z}+\gamma \hat{\tau}^{x},
\end{eqnarray}
where $\hat{\tau}^{\mu} (\mu=x,y,z)$ is Pauli matrix in the orbital basis $(f_{1},f_{2})$, and
$\hat{\tau}^{0}$ is identity matrix. In the presence of $g$, the Coulomb interaction $u\hat{U}^{0;c}$
is replaced with $u\hat{U}^{0;c}+2g\hat{W}$.
In the present numerical study, we put $(\alpha,\beta,\gamma)=(0,1,-1)$.
We verified that the main results are qualitatively 
same as those of $(\alpha,\beta,\gamma)=(0,1,1)$.
The numerical results are not sensitive to the ratio of $(\alpha,\beta,\gamma)$.

This effective interaction can be induced by (for instance) 
the electron-phonon interaction due to 
$A_{1}^{+}$ mode, such as
the oscillation of $c$-axis length \cite{Kontani-RPA}.
 In this case, $g$ is expressed as
 $g=\tilde{g}\frac{\omega_{D}^2}{\omega_{D}^{2}+\omega_{j}^{2}}$, where 
 $\tilde{g}=\frac{2\eta^{2}}{\omega_{D}}\ (>0)$: $\omega_{D}$ is the phonon frequency,
$\eta$ is the coupling constant between electrons and phonon, and
 $\omega_{j}=2j\pi T$ is the Boson Matsubara frequency.
 In the present study, we drop $\omega_{j}$-dependence of $g$ 
for simplicity. That is, we neglect the retardation effect,
which leads to underestimation of the $s$-wave 
superconducting $T_{c}$ as discussed in Ref.\cite{rina2}.
The $A^{+}_{1}$ effective interaction in Eq. (\ref{eqn:ph}) is classified into
even-rank multipole interaction.
Therefore, strong electric (=even-rank) multipole fluctuations are induced by the interaction $g$.
On the other hand, the magnetic (=odd-rank) multipole susceptibilities are
 independent of $g$. 

\section{Green function}
\label{sec:Green}
Here, we introduce the Green functions in the present model.
The $3\times 3$ matrix form of the Green functions is given by
\begin{eqnarray}
\hat{G}^{\sigma}(\k,i\epsilon_{n})=\left((i\epsilon_{n}-\mu)\hat{1}-\hat{h}_{\k}^{\sigma} \right)^{-1},
\label{eqn:Green1}
\end{eqnarray}
where $\hat{h}_{\k}^{\sigma}$ is introduced in Eq. (\ref{eqn:hamiltonian2}).
The first two rows and columns of Eq. (\ref{eqn:Green1})
give the $f$-orbital  Green functions. They are expressed as
 \begin{eqnarray}
G^{f,\sigma}_{lm}\!(k)\!=\!G^{0f}_{l}\!(k)\delta_{lm}\!\!+\!\!G^{0f}_{l}\!(k)V^{*}_{\k l\sigma}G^{c,\sigma}\!(k)V_{\k m\sigma}G^{0f}_{m}\!(k),
\label{eqn:Green2}
\end{eqnarray} 
where $l,m=1,2$, $k=(\k,\e_n)=(\k,(2n+1)\pi T)$, and
\begin{eqnarray}
G^{0f}_{l}(k)=\left(i\epsilon_{n}-\mu-E_{l} \right)^{-1}.
\label{eqn:Gree3}
\end{eqnarray} 
$G^{c,\sigma}(\k)$ is the $s$-electron Green function given by the (3,3) component 
of Eq. (\ref{eqn:Green1}). It is expressed as
\begin{eqnarray}
G^{c,\sigma}(k)=\left(i\epsilon_{n}-\mu-\epsilon_{\k}-\sum_{l} V_{\k l\sigma}G^{0f}_{l}(k)V^{*}_{\k l\sigma} \right)^{-1}\!\!\!\!.
\label{eqn:Gree4}
\end{eqnarray} 
In the present two-orbital model, the relation
$V^{*}_{\k l\uparrow}V_{\k m\uparrow}=V^{*}_{\k l\downarrow}V_{\k m\downarrow}$
is satisfied, as we can verify from Eq. (\ref{eqn:hopping}).
For this reason, the Green functions become 
independent of spin index:
\begin{eqnarray}
G^{f}_{lm}(k)\equiv G^{f,\uparrow}_{lm}(k)=G^{f,\downarrow}_{lm}(k),\nonumber \\
G^{c}(k)\equiv G^{c,\uparrow}(k)=G^{c,\downarrow}(k).\end{eqnarray} 
In the present model, diagonal ($l=m$) components of 
$G^{f}_{lm}(k)$ and off-diagonal ($l\neq m$) ones
 are comparable since each $s$-$f$ hybridization 
in Eq. (\ref{eqn:hopping}) is isotropic in magnitude.
 It is a characteristic feature of the multiorbital $f$-electron systems.
\section{Susceptibility}
\label{sec:RPA}
\begin{figure}[htb]
\includegraphics[width=.90\linewidth]{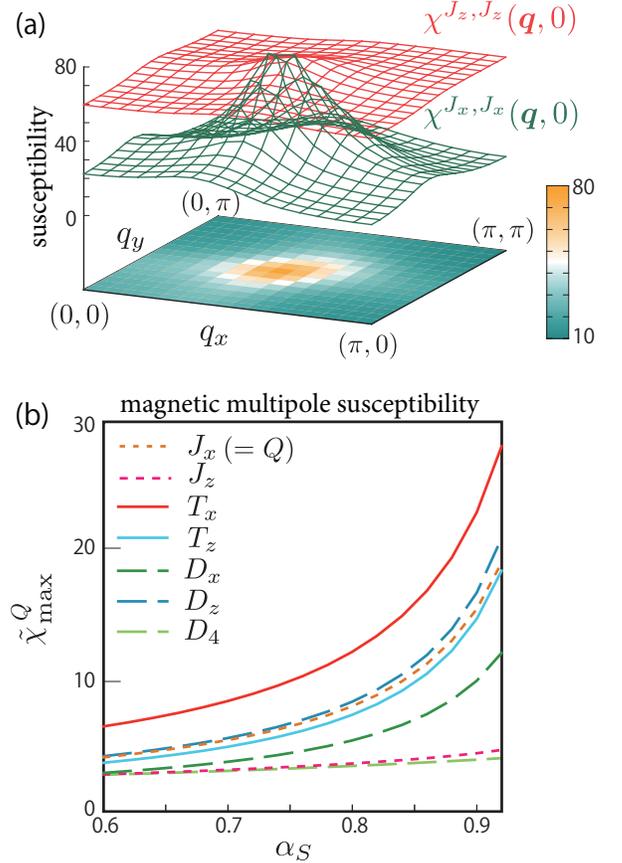}
\caption{(a) $\q$ dependence of the magnetic dipole susceptibility.
$\chi^{J_{z},J_{z}}(\q,0)\gg \chi^{J_{x},J_{x}}(\q,0)$ is satisfied at $\q=(0,0)$.
(b) $\a_{S}$ dependence of magnetic multipole susceptibility.
Higher-rank magnetic multipole
 susceptibilities are strongly enlarged.}
\label{fig:chi}
\end{figure}
First, we perform the random phase approximation (RPA) 
in order to obtain the $f$-electron susceptibility. 
In this calculation, we use $32\times 32$ $\k$-meshes 
and $128$ Matsubara frequencies. 
The irreducible susceptibility of $f$-electron is given by
\begin{eqnarray}
\chi_{ll'mm'}^{0}(q)= -T\sum_{k}G^{f}_{l m}(k+q)G^{f}_{m' l'}(k),
\label{eqn:chi0}
\end{eqnarray}
where $q=(\q,\w_j)=(\q,2j\pi T)$.
In the RPA, the susceptibility for each channel ($ch$) is given as
\begin{eqnarray}
\hat{\chi}^{ch}(q)= {\hat \chi}^{0}(q)
({\hat 1}-u{\hat U}^{0;ch}{\hat \chi}^{0}(q))^{-1} ,
 \label{eqn:chi}
\end{eqnarray}
where $\hat{\chi}^{ch}(q)$ is $2^{2}\times 2^{2}$ matrix. 
Using $\hat{\chi}^{ch}$ ($ch=s,s\!\! \perp,c$), the $f$-electron susceptibility in the $L=(l,\sigma)$
 basis is expressed as
\begin{eqnarray}
\hat{\chi}^{\sigma \sigma' \lambda \lambda'}
=\frac{1}{2}\hat{\chi}^{s}
(\sigma^{x}_{\sigma \sigma'} \sigma^{x}_{\lambda' \lambda}+
\sigma^{y}_{\sigma \sigma'} \sigma^{y}_{\lambda' \lambda})\nonumber \\
+\frac{1}{2}\hat{\chi}^{s\perp}
\sigma^{z}_{\sigma \sigma'} \sigma^{z}_{\lambda' \lambda}
+\frac{1}{2}\hat{\chi}^{c} \sigma^{0}_{\sigma \sigma'}\sigma^{0}_{\lambda' \lambda}.
 \label{eqn:chisc}
\end{eqnarray}
Here, we define the pseudo-spin Stoner factor $\alpha_{S} (\alpha_{S\perp})$ 
as the largest eigenvalue of $u{\hat U}^{0;s(s\perp)}{\hat \chi}^{0}(q)$.
In the present model, each matrix element of $\hat{U}^{0;s}$ and that of $\hat{U}^{0;s\perp}$
in TABLE \ref{tab:coulomb} are the same except for $(lmlm)$ and $(llmm)$ elements.
For this reason,  $\hat{\chi}^{s}\approx \hat{\chi}^{s\perp}$ 
and $\alpha_{S} \approx \alpha_{S\perp}$ are satisfied.

Now, we define the multipole susceptibility for $Q$$(=(\Gamma,\phi))$;
\begin{eqnarray}
\chi^{Q,Q'}(q)=\int^{\beta}_{0} d\tau \left\langle {\cal O}^{Q}({\bm q},\tau){\cal O}^{Q'}({\bm -\q},\tau)\right\rangle e^{i\w_j\tau},
\label{eq:suscep}
\end{eqnarray} where ${\cal O}^{Q}(\q, \tau)=\sum_{L,M,\k} O^{Q}_{L,M} 
f^{\dagger}_{\k m\sigma}(\tau) f_{\k +\q l \sigma'}(\tau)$.
In 3D models, $\chi^{(\Gamma,\phi),(\Gamma',\phi')}(q)$ can be finite 
even in the case of $\Gamma \neq \Gamma'$.
In contrast, in the present 2D model,  $\chi^{(\Gamma,\phi),(\Gamma',\phi')}(q)=0$ for any $q$ in the case of $\Gamma \neq \Gamma'$, which is a great merit of the present model in analysis.
Note that $\chi^{(\Gamma,\phi),(\Gamma,\phi')}(q)$ for $\Gamma=A^{+}_{1},A^{+}_{2},E^{+}$ is
 classified into electric susceptibility, and that
 for $\Gamma=A^{-}_{1},A^{-}_{2},E^{-}$ is classified into magnetic susceptibility.
In the present model, $\alpha_{S}$ corresponds to the $A^{-}_{2}$ 
magnetic (=odd-rank) susceptibility, that is, $\alpha_{S}=\alpha_{A^{-}_{2}}$ in the RPA.
We obtain the relation $1 \gtrsim \alpha_{A^{-}_{2}} \gtrsim \alpha_{E^{-}}$.

In Fig.\ref{fig:chi}(a), we show obtained susceptibilities at $u=0.31$
for the magnetic dipole 
$J_{z} =(A^{-}_{2},1)$, $\chi^{J_{z},J_{z}}(\q,0)$, and
$J_{x}=(E^{-},1)$, $\chi^{J_{x},J_{x}}(\q ,0)$.
In this case, $\alpha_{S}=0.90$.
Note that 
$\chi^{J_{x},J_{x}}=\chi^{J_{y},J_{y}}$. 
We find that $\chi^{J_{z},J_{z}}(\q,0)$ is 
much larger than $\chi^{J_{x},J_{x}}(\q,0)$ at $\q=(0,0)$
while they are almost the same around the peak at $\q\simeq (\pi/2,\pi/2)$.
Thus, the uniform magnetic susceptibility shows strong Ising anisotropy, which is
actually observed in CeCu$_{2}$Si$_{2}$.

Hereafter, to compare among different-rank of multipole susceptibilities, 
we define normalized multipole operator $\hat{\tilde{O}}^{Q}$ 
as $\mathrm{Tr}(\hat{\tilde{O}}^{Q2})=1$, that is, 
\begin{eqnarray}\hat{\tilde{O}}^{Q}=\hat{O}^{Q}/\sqrt{\mathrm{Tr} (\hat{O}^{Q2})}.
\label{eqn:normal}
\end{eqnarray}
The normalized susceptibility $\tilde{\chi}^{Q,Q'}(q)$ is given
 by replacing $\hat{O}^{Q}$ in Eq. (\ref{eq:suscep}) with $\hat{\tilde{O}}^{Q}$.
In Fig.\ref{fig:chi}(b), we show $\a_{S}$ dependences of the maximum of magnetic multipole susceptibilities
$\tilde{\chi}_{\rm{max}}^{Q}\equiv \max_{\q} \{ \tilde{\chi}^{Q,Q}(\q,0)\}$. 
$\alpha_{S}$ linearly increases in proportion to $u$.
The obtained $\tilde{\chi}_{\rm{max}}^{Q}$ is the most divergent for $Q=T_{x}$.
This fact is consistent with the RPA result based on the first-principles model in Ref\cite{Ikeda-122}.
Secondly, $\tilde{\chi}_{\rm{max}}^{Q}$ for $Q=D_{z},J_{x},T_{z},D_{4}$ is also strongly enlarged.
Therefore, various magnetic multipole (including higher-rank) susceptibilities are
simultaneously enlarged in the RPA.
This is a characteristic feature of $f$-electron systems with strong SOI \cite{Ikeda-Uru2}.
We find that the inter-rank magnetic multipole susceptibilities, such as $\tilde{\chi}^{J_{z},T_{z}}$, are also enlarged.

Now, we explain the reason why higher-rank magnetic multipole susceptibilities are enlarged.
Our result means that orbital-off-diagonal components of 
$\chi^{s}_{ll'mm'}$ are comparable to orbital-diagonal ones. 
In fact, $\chi^{s}_{1111}\approx \chi^{s}_{1112}$ is satisfied in the present model. 
It originates from the fact that each $s$-$f$ hybridization in Eq. (\ref{eqn:hopping})
is isotropic in the $x$- and $y$-directions due to the strong SOI, and therefore
each $f_{l}$-orbital weight is independent of
$\theta$ as shown in Fig.\ref{fig:band}(e).
This is the origin of the large orbital-off-diagonal components of
$G_{lm}^{f}$ and those of $\chi_{ll'mm'}^{s(s\perp)}$.
This situation is quite different from 3$d$-electron systems, in which off-diagonal components 
of $\hat{G}$ and $\hat{\chi}^{s}$ remain small in general. 

Finally, we comment on electric (=even-rank) susceptibilities obtained by the RPA.
In the absence of electric multipole-multipole (phonon-induced) interaction: $g=0$,
 the obtained electric susceptibilities are much smaller than magnetic ones.
 That is, charge stoner factor $\alpha_{C}$, which is defined as the largest eigenvalue of 
 $(u\hat{U}^{0;c}+2g\hat{W})\hat{\chi}^{0}(q)$, satisfy $\alpha_{C} \ll \alpha_{S}$.
 In the present model, $(\alpha_{C},\alpha_{S})=(0.55,0.90)$.
 On the other hand, 
 the hexadecapole $\chi^{H_{0},H_{0}}$ and quadrupole 
 $\chi^{O_{20},O_{20}}$ susceptibility are enlarged at $q\approx (\pi,\pi)$
 when we consider the small $g$ $(>0)$ .
In this case, $\alpha_{C}$ increases to $0.84$ at $g=0.04$. 
Note that the obtained electric susceptibilities work as attractive interaction for
$s$-wave superconductivity, as we will explain in the following section.

In principle, some experimental signatures due to the electric multipole susceptibility
may be observed. For instance, enhancement of $\chi^{c}$ at $q =0$ can induce
the softening of elastic constants.
Also, enhancement of $\chi^{c}$ at $q \neq 0$ may be observed by neutron scattering experiment as softening of phonon dispersion.
\section{Gap equation}
\label{sec:SC}
Now, we solve the linearized gap equation by focusing on the important roles of the vertex corrections, which we call $U$-VC.
The bare electron-boson couplings are dressed by the $U$-VC, which is totally  dropped in conventional Migdal approximation.
The gap equation for spin-singlet paring in the band basis is given as
\begin{eqnarray}
\lambda \Delta(\k,\epsilon_{n})=
-\frac{\pi T}{(2\pi)^2}\sum_{\epsilon_{m}}
\oint_{FS} \frac{d\k'}{v_{\k'}} 
\frac{\Delta(\k',\epsilon_{m})}{|\epsilon_{m}|} V^{\rm{sing}}_{k,k'} ,\label{eqn:linear}
\end{eqnarray}
where $\Delta(\k,\epsilon_{n})$ is the gap function on Fermi surface,
$\lambda$ is the eigenvalue, and
$v_{\k}$ is the Fermi velocity on Fermi surface.
${V}^{\rm{sing}}_{k,k'}$ is the 
spin singlet paring interaction including $U$-VC.
The diagrammatic expression of the gap equation is shown in Fig.\ref{fig:diagram}(a).
The black triangle shows the three-point vertex correction due to many body effects.
We consider the AL-type diagram for $U$-VC given in Fig.\ref{fig:diagram}(b),
 which is explained in more detail in the Section \ref{sec:U-VC}.
The paring interaction in Eq. (\ref{eqn:linear}) is obtained by
\begin{eqnarray}
V^{\rm{sing}}_{k,k'}&=&V^{udud}_{k,k'}-
V^{uudd}_{k,k'} \nonumber \\
&=&\frac{1}{2}\sum_{\Sigma,\Lambda}V^{\Sigma \Lambda \bar{\Lambda} \bar{\Sigma}}_{k,k'}
(1-2\delta_{\Sigma\Lambda}),
\label{eqn:linear2}
\end{eqnarray}
where $\Sigma,\Lambda=u\,\,(d)$ is pseudo-spins up (down) 
that denotes the Kramers doublet of the Bloch function, and $\bar{\Sigma}\equiv -\Sigma$.
$V^{\Sigma \Lambda \bar{\Lambda} \bar{\Sigma}}_{k,k'}$ is given as
\begin{eqnarray}
V^{\Sigma \Lambda \bar{\Lambda} \bar{\Sigma}}_{k,k'}=
\sum_{ll'mm'} \sum_{\sigma\sigma'\lambda\lambda'}
U^{\Sigma*}_{l\sigma}(\k)
U^{\bar{\Sigma}*}_{m'\lambda'}(-\k)  \nonumber \\ 
\times V^{\sigma \sigma' \lambda \lambda'}_{ll'mm'}(k,k')
U^{\bar{\Lambda}}_{m\lambda}(-\k')
U^{\Lambda}_{l'\sigma'}(\k')\label{eqn:unitary},
\end{eqnarray}
where $U^{\Sigma}_{l\sigma}(\k)$ is the unitary matrix connecting between
$f^{\dagger}_{\k l \sigma}$ and the quasi-particle creation operator $f^{\dagger}_{\k \Sigma}$.
In the presence of the time reversal symmetry, $U^{\Sigma}_{l\sigma}(\k)$ is
 related to $U^{\bar{\Sigma}}_{l\bar{\sigma}}(\k)$ as
$U^{\Sigma}_{l\sigma}(-\k)=(-1)^{\delta_{\Sigma \sigma}+1}U^{\bar{\Sigma}}_{l\bar{\sigma}}(\k)^{*}$  
\cite{saito}.
\begin{figure}[htb]
\includegraphics[width=.96\linewidth]{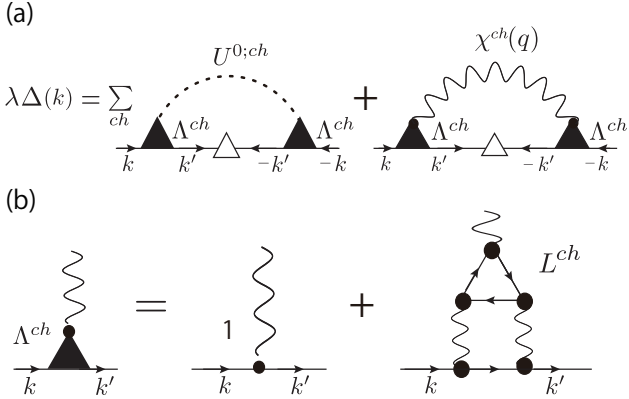}
\caption{(a) Linearized gap equation in the present study.
 The black triangle shows three-point vertex correction ($U$-VC). 
(b) $U$-VC due to the AL process.}
\label{fig:diagram}
\end{figure}
$V^{\sigma \sigma' \lambda \lambda'}_{ll'mm'}(k,k')$ is the paring interaction in the orbital basis
introduced in Sec. \ref{sec:U-VC}.
In the present model, there is rotational symmetry along $z$-axis in the pseudo-spin space.
For this reason, $V^{\sigma \sigma' \lambda \lambda'}_{ll'mm'}(k,k')$
is uniquely decomposed into spin and charge channels as follows
\begin{eqnarray}
V^{\sigma \sigma' \lambda \lambda'}_{ll'mm'}=
\frac{1}{2}V^{s\perp}_{ll'mm'}
(\sigma^{x}_{\sigma \sigma'} \sigma^{x}_{\lambda' \lambda}+
\sigma^{y}_{\sigma \sigma'} \sigma^{y}_{\lambda' \lambda})\nonumber \\
+\frac{1}{2}V^{s}_{ll'mm'}
\sigma^{z}_{\sigma \sigma'} \sigma^{z}_{\lambda' \lambda}
+\frac{1}{2}V^{c}_{ll'mm'}\sigma^{0}_{\sigma \sigma'}\sigma^{0}_{\lambda' \lambda},
\label{eqn:Vparaperp}
\end{eqnarray}
where we drop the first order term ($U^{0;s\perp}$) from $V^{0;s\perp}$
in order to avoid double counting \cite{Yamakawa-FeSe2}.
From Eqs.(\ref{eqn:linear2})-(\ref{eqn:Vparaperp}),
we obtain that 
\begin{eqnarray}
V^{\rm{sing}}_{k,k'}&\hspace{-7pt}=\hspace{-7pt}&\sum_{ll'mm'}
V^{s\perp}_{ll'mm'}
\left(\hat{A}^{udud}_{\uparrow\downarrow\uparrow\downarrow}-
\hat{A}^{uddu}_{\uparrow\downarrow\uparrow\downarrow}
\right)_{ll'mm'}\nonumber \\ 
&\hspace{-7pt}+\hspace{-7pt}&\frac{1}{2} V^{c}_{ll'mm'} \left(
\hat{A}^{udud}_{\uparrow\uparrow\uparrow\uparrow}+
\hat{A}^{udud}_{\uparrow\downarrow\downarrow\uparrow}-
\hat{A}^{uddu}_{\uparrow\uparrow\uparrow\uparrow}-
\hat{A}^{uddu}_{\uparrow\downarrow\downarrow\uparrow}
\right)_{ll'mm'} \nonumber \\ 
&\hspace{-7pt}+\hspace{-7pt}&
\frac{1}{2}V^{s}_{ll'mm'}\left(
\hat{A}^{udud}_{\uparrow\uparrow\uparrow\uparrow}-
\hat{A}^{udud}_{\uparrow\downarrow\downarrow\uparrow}-
\hat{A}^{uddu}_{\uparrow\uparrow\uparrow\uparrow}+
\hat{A}^{uddu}_{\uparrow\downarrow\downarrow\uparrow}
\right)_{ll'mm'},
\label{eqn:Vsc}
\end{eqnarray} where 
\begin{eqnarray}
(\hat{A}^{\Sigma \bar{\Sigma} \bar{\Lambda}  \Lambda }_{\sigma \sigma' \lambda \lambda'})_{ll'mm'}
&\equiv & U^{\Sigma*}_{l\sigma}(\k)
U^{\bar{\Sigma} *}_{m'\sigma'}(-\k)
U^{\bar{\Lambda}}_{m\lambda}(-\k')
U^{\Lambda}_{l'\lambda'}(\k')\label{eqn:Adet}.\nonumber
\end{eqnarray}
In the present model, the electric multipole paring interaction corresponds to
attraction, while the magnetic one corresponds to
repulsion.
To understand this fact, we consider the paring interaction 
in the absence of SOI, like in 3$d$-electron systems.
In this case, we can put $U^{\Sigma*}_{l\sigma}(\k)=U^{*}_{l}(\k)\delta_{\Sigma,\sigma}$
and 
\begin{eqnarray*}
\begin{cases}\hat{A}^{udud}_{\uparrow\downarrow\uparrow\downarrow}=
\hat{A}^{uddu}_{\uparrow\downarrow\downarrow\uparrow}\neq 0
 \\ 
\hat{A}^{uddu}_{\uparrow\downarrow\uparrow\downarrow}=
\hat{A}^{udud}_{\uparrow\uparrow\uparrow\uparrow}=
\hat{A}^{udud}_{\uparrow\downarrow\downarrow\uparrow}=
\hat{A}^{uddu}_{\uparrow\uparrow\uparrow\uparrow}=0.\end{cases}
\label{eqn:delta}
\end{eqnarray*}
Then, we obtain the following simple expression:
\begin{eqnarray*}
V^{\rm{no\mbox{-}SOI}}_{k,k'}\hspace{-2pt}=\hspace{-7pt}\sum_{ll'mm'}(\hat{V}^{s\perp}+\frac{1}{2}\hat{V}^{s}-\frac{1}{2}\hat{V}^{c})_{ll'mm'}(\hat{A}^{udud}_{\uparrow\downarrow\uparrow\downarrow})_{ll'mm'},
\label{eqn:Vsingeasy}
\end{eqnarray*}
where $V^{s\perp}=V^{s}$ is satisfied when SOI is dropped.
Thus, $V^{\rm{sing}}_{k,k'}$ in Eq. (\ref{eqn:Vsc}) is reduced to the well-known 
expression $V^{\rm{no\mbox{-}SOI}}_{k,k'}\propto \frac{3}{2}V^{s}-\frac{1}{2}V^{c}$.
In conclusion, the charge- or electric-channel paring interaction works as attraction,
while the spin- or magnetic-channel one works as repulsion. 

\section{Important roles of $U$-VC}
\label{sec:U-VC}
Here, we discuss about the important roles of 
$U$-VC in the paring interaction.
Until now, $U$-VC in $d$-electron systems has been studied intensively 
by some theoretical methods, such as the functional renormalization group (fRG) theory
 and perturbation theory.
Both theoretical frameworks reveal that
$U$-VC makes significant contribution to the superconductivity,
especially in multi-orbital systems, so
Migdal approximation fails.
In more detail, AL-type $U$-VC becomes more important 
than MT-type one near the magnetic QCP. 
However, $U$-VC in $f$-electron system with strong SOI has not been understood at all.
In the present study, we show that $U$-VC in $f$-electron systems is
more important than that in $d$-electron systems due to large SOI.

Now, we discuss about the paring interaction with $U$-VC.
In the present model, $U$-VC satisfy the pseudo-spin conservation.
Thus, the paring interaction for each channel
in Eq. (\ref{eqn:Vsc}) is expressed as
\begin{eqnarray}
\hat{V}^{ch}(k,k')={\hat \Lambda}^{ch}_{k,k'}
{\hat I}^{ch}(k-k'){\hat {\bar \Lambda}}^{ch}_{-k,-k'},
\label{eqn:Vch}
\end{eqnarray}where\vspace{-5pt}
\begin{eqnarray}
{\hat I}^{ch}(k-k')=u^{2}{\hat U}^{0;ch}{\hat \chi}^{ch}(k-k'){\hat U}^{0;ch}+u{\hat U}^{0;ch}.
\label{eqn:Vchi}
\end{eqnarray}
Here, ${\hat \Lambda}^{ch}_{k,k'}$ is an enhancement factor for electron-boson coupling given by 
\begin{eqnarray}( {\hat \Lambda}^{ch}_{k,k'})_{ll'mm'} = \delta_{lm}\delta_{l'm'}+
({\hat L}^{ch}_{k,k'})_{ll'mm'},
\label{eqn:defALc}
\end{eqnarray} where ${\hat L}^{ch}_{k,k'}$ is AL-type $U$-VC,
whose diagrammatic expression is given in Fig.\ref{fig:diagram}(b).
In Eq. (\ref{eqn:Vch}), $({\hat {\bar \Lambda}}^{ch}_{k,k'})_{ll'mm'}\equiv
({\hat \Lambda}^{ch}_{k,k'})_{m'ml'l}$.
In the present model, the MT-type $U$-VC is negligible compared to AL-type one.
For this reason, we calculate only AL-type $U$-VC.
Note that $\hat{V}^{ch}=\hat{I}^{ch}$ in the Migdal approximation ($\hat{\Lambda}^{ch}=\hat{1}$).

Hereafter, we discuss only the charge-channel $U$-VC ${\hat \Lambda}^{c}_{k,k'}$
 since it becomes much larger than unity near the magnetic
  QCP, whereas spin-channel one remains order of unity.
  Hence, the charge-channel paring interaction is enlarged by $\hat{\Lambda}^{c}_{k,k'}$.
Here, $\hat{L}^{c}_{k,k'}$ is derived from the $U$-VC in the $(l,\sigma)$ basis.
\begin{eqnarray}
\hat{L}^{c}_{k,k'} \equiv 
\hat{L}_{k,k'}^{\uparrow\uparrow\uparrow\uparrow}\!\!+\hat{L}_{k,k'}^{\uparrow\uparrow\downarrow\downarrow}
=\hat{L}_{k,k'}^{\downarrow\downarrow\downarrow\downarrow}\!\!+\hat{L}_{k,k'}^{\downarrow\downarrow\uparrow\uparrow},
\label{eqn:lamcdef}
\end{eqnarray}
whose Feynman diagram is shown in Fig.\ref{fig:uvchitai1}(a). 
The analytic expression of $\hat{L}^{c}_{k,k'}$ is given as
\begin{eqnarray}
\!\!(\hat{L}^{c}_{k,k'})_{ll'mm'}\!\!\!\!\!\!&&=\!\frac{T}{2}\!\!\!\sum_{p,abcdef}
B_{abcdef}^{mm'} (k-k',p) \nonumber \\ 
&&\times \sum_{ch} a^{ch} I^{ch}_{lacd} (k-k'+p) I^{ch}_{bl'ef} (-p),
\label{eqn:UALc}
\end{eqnarray}
where $(a^{s},a^{s\perp}a^{c})=(1,2,1)$, and
\begin{eqnarray}
&& \hspace{-10pt} B_{abcdef}^{mm'} (q,p)\!\equiv \!\frac{1}{4}G^{f}_{ab}(k'\!\!\!-p)\!\left\{\! C^{mm'}_{cdef} (q,p)\!+\!
C^{mm'}_{efcd}(q,q+p)\!\right\}\!\!, \nonumber \\ &&  
\label{eqn:Bdef3}
\end{eqnarray}\vspace{-9pt}
\begin{eqnarray}
C^{ab}_{cdef} (q,p)\!\equiv \!-T\sum_{k'}G^{f}_{ca}(k'+q)G^{f}_{bf}(k')G^{f}_{ed}(k'-p).
\end{eqnarray}
Here, $a\sim f$ are orbital indices. 
In the present numerical study, 
we put $g=0$ in the $\hat{L}^{ch}_{k,k'}$, since
the contribution from $\chi^{c}$
is negligibly smaller than that from $\chi^{s}$ and $\chi^{s\perp}$ \cite{rina2}.

Next, we show numerical results of $\hat{\Lambda}^{c}_{k,k'}$.
Here, we use $16\times16$ $\k$-meshes and 128 Matsubara frequencies.
In Figs.\ref{fig:uvchitai1}(b) and \ref{fig:uvchitai1}(c), we show the $\alpha_{S}$ dependence of  
maximum value of ${\hat \Lambda}^{c}_{k,k'}$ on the Fermi surface,
\begin{eqnarray}
\Lambda^{c,{\rm max}}_{ll'mm'}\equiv \max_{\k,\k' \in FS}| (\hat{\Lambda}^{c}_{k,k'})_{ll'mm'}|,
\label{eqn:fig4bc}
\end{eqnarray}
at $\epsilon_{n}=\epsilon_{n'}=\pi T$. 
We plot various orbital components of $U$-VC.
Note that the other elements are obtained by using the symmetry relation
of orbital indices, that is,
$\Lambda^{c,{\rm max}}_{ll'mm'}=\Lambda^{c,{\rm max}}_{l'lm'm}$.
We find that they work as large enhancement factors for the
coupling constant between electrons and charged-bosons ($|\hat{\Lambda}^{c}| \gg 1$)
near the magnetic QCP ($\alpha_{S}\lesssim 1$). 
Note that all magnetic multipole susceptibilities except for $D_{4}$, $Q=(A^{-}_{1},1)$, are included
in either $\chi^{s}$ or $\chi^{s\perp}$.
This behavior originates from the relation
$\hat{\Lambda}^{c}_{k,k'}\propto 
\sum_{p} \hat{\chi}^{s}(k-k'+p) \hat{\chi}^{s}(p) +2 \hat{\chi}^{s\perp}(k-k'+p) \hat{\chi}^{s\perp}(p)$.
This is qualitatively similar to $d$-electron systems 
without SOI as shown in Fig.2(c) in Ref.\cite{rina2}.
In conclusion, $U$-VC in $f$-electron systems 
give significant contribution as well as
in $d$-electron systems.
\begin{figure}[htb]
\includegraphics[width=.92\linewidth]{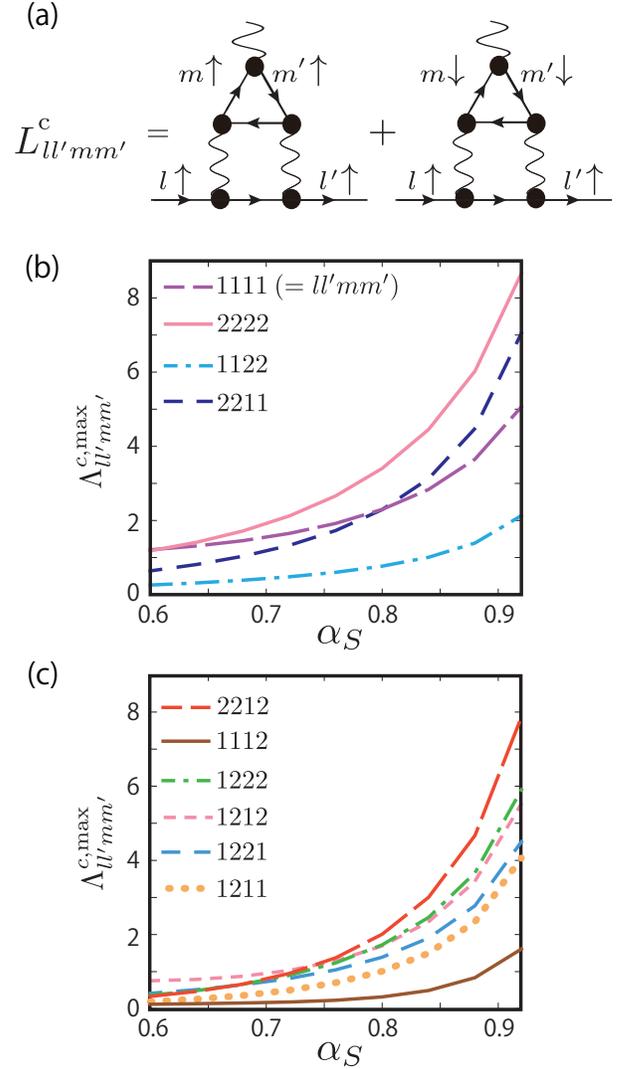}
\caption{(a) Charge-channel $U$-VC given by AL process. 
Only the diagrams given by the first term of $\hat{B}$ in Eq. (\ref{eqn:Bdef3}) are shown.
(b) and (c) $\alpha_{S}$ dependence of  charge-channel $U$-VC
$\Lambda_{ll'mm'}^{c,\rm{max}}$. Various orbital components are strongly enlarged. }
\label{fig:uvchitai1}
\end{figure}

We stress that there are some significant differences from $d$-electron systems.
In fact, in the present $f$-electron system, 
(i) various orbital components of $U$-VC are equally enlarged,
and (ii) the magnitude of $U$-VC are even larger than in $d$-electron systems at the same $\alpha_{S}$. 
These results originate from multiple (higher-rank) 
magnetic multipole fluctuations as shown in Fig.\ref{fig:chi}(b).
To clarify this fact, we are going to elucidate what types of
multipole fluctuations are significant for $U$-VC below.
We recall that the $f$-electron susceptibility in Eq. (\ref{eqn:chisc}) is uniquely 
expanded on the basis of $4\times4$ matrix expression of multipole operator
$\hat{O}^{Q}(=\hat{O}^{(\Gamma,\phi)})$ given in Appendix C as follows,
\begin{eqnarray}
\chi_{LL'MM'}(q)&=&\sum_{\Gamma,\phi,\phi'}
a^{\Gamma,\phi,\phi'}(q)
O_{LL'}^{(\Gamma,\phi)}
O_{MM'}^{(\Gamma,\phi')*}.
\label{eqn:shaei2}
\end{eqnarray}
Note that $\sum_{LL'}O_{LL'}^{(\Gamma,\phi)}O_{LL'}^{(\Gamma,\phi')*}=0$ for
$\Gamma \neq \Gamma'$.
The derivation of the coefficient $a^{\Gamma,\phi,\phi'}(q)$ is explained in Appendix D.
In the same way, the interaction $\hat{I}(=u^{2}\hat{U}^{0}\hat{\chi}^{ch}\hat{U}^{0}+u\hat{U}^{0})$ in the $L=(l,\sigma)$ basis is expanded as
\begin{eqnarray}
I_{LL'MM'}&=&\sum_{\Gamma,\phi,\phi'}b^{\Gamma,\phi,\phi'}(q)
O_{LL'}^{(\Gamma,\phi)}
O_{MM'}^{(\Gamma,\phi')*}.\label{eqn:shaei3}
\end{eqnarray}
%
By utilizing the pseudo-spin conservation law, each term in the right-hand-side
of Eq. (\ref{eqn:shaei3}) is expressed in the $l$-basis as $I^{ch,\Gamma, \phi, \phi'}_{ll'mm'}$. Note that $\hat{I}^{ch,\Gamma, \phi, \phi'}=0$ for $ch \neq ch_{\Gamma}$
By replacing $I^{ch}_{ll'mm'}$ in Eq. (\ref{eqn:UALc}) with 
$I^{ch,Q}_{ll'mm'}$$(\equiv I^{ch,\Gamma,\phi,\phi}_{ll'mm'})$,
we obtain multipole-decomposed $U$-VC symbolically expressed as
\begin{eqnarray}
(\hat{L}^{c, QQ}_{k,k'})&=&\frac{T}{2} \sum_{ch}
\hat{B} \hat{I}^{ch,Q} \hat{I}^{ch,Q},\label{eqn:projQ} \\
(\hat{L}^{c, QQ'}_{k,k'})&=&\frac{T}{2} \sum_{ch}
\hat{B} \left( \hat{I}^{ch,Q} \hat{I}^{ch,Q'} + \hat{I}^{ch,Q'} \hat{I}^{ch,Q} \right).
\label{eqn:projQQ}
\end{eqnarray} where $Q\neq Q'$. The diagrammatic expression of 
Eq. (\ref{eqn:projQQ}) is given in Fig.\ref{fig:uvchitai2}(a). 
Note that the relation $\hat{L}^{c} \approx \sum_{\{Q,Q'\}}\hat{L}^{c,QQ'}$ is satisfied.
${\hat \Lambda}^{ch QQ'}$ is given by
\begin{eqnarray}( {\hat \Lambda}^{ch,QQ'}_{k,k'})_{ll'mm'} = \delta_{lm}\delta_{l'm'}+
({\hat L}^{ch,QQ'}_{k,k'})_{ll'mm'}
\label{eqn:defALcQQ}
\end{eqnarray}

In Figs.\ref{fig:uvchitai2}(b)-(e), we show the maximum
of multipole-decomposed $U$-VC defined as 
\begin{eqnarray}
\Lambda^{c,QQ'}_{ll'mm'} \equiv \max_{\k,\k' \in FS} |(\hat{\Lambda}^{c,QQ'}_{k,k'})_{ll'mm'}|
\label{eqn:projQQ2}
\end{eqnarray} at $\epsilon_{n}=\epsilon_{n'}=\pi T$.
We consider only odd-rank (=magnetic) multipole operators for $Q$ and $Q'$ since
 the contributions from even-rank multipole operators are negligibly small in RPA.
In addition, $\hat{\Lambda}^{c,QQ'}_{ll'mm'}$ with 
$Q=(\Gamma,\phi)$ and
$Q'=(\Gamma',\phi')$ becomes zero
except for $\Gamma=\Gamma'$ in the present model.
Figures \ref{fig:uvchitai2}(b) and (c) show the orbital-diagonal component of $U$-VC
given by $\Lambda^{c,QQ'}_{2222}$.
It becomes the largest for $(Q,Q')=(T_{x},T_{x})$.
Subsequently, $(Q,Q')=(J_{z},T_{z}), (T_{z},T_{z}), (D_{z},D_{z})$ are also enlarged.
In Figs.\ref{fig:uvchitai2}(d) and (e), we show orbital-off-diagonal component given by
$\Lambda^{c,QQ'}_{1211}$.
It takes the largest value for $(Q,Q')=(T_{x},D_{x})$.
Its value for $(Q,Q')=(T_{z},D_{z}), (D_{z},D_{z}), (T_{x},T_{x}),(J_{z},D_{z})$ are also enlarged.

In summary, in heavy fermion systems, multiple multipole fluctuations lead to the
strong enhancement of $U$-VC, $\Lambda^{c}$.
In Figs.\ref{fig:uvchitai1}(b) and (c),
both orbital-diagonal and off-diagonal components of $\Lambda^{c}$  are enlarged.
In Figs.\ref{fig:uvchitai2}(b)-(e), many pairs of multipole fluctuations (Q,Q') contribute
to the enhancement of $\Lambda^{c}$.
These facts lead to above-mentioned differences (i) and (ii), which are not seen in 3$d$-electron system. 
Thus, we conclude that the $U$-VC in $f$-electron system plays more significant roles due to the strong SOI compared to $3d$-electron systems.  
\begin{figure}[htb]
\includegraphics[width=.94\linewidth]{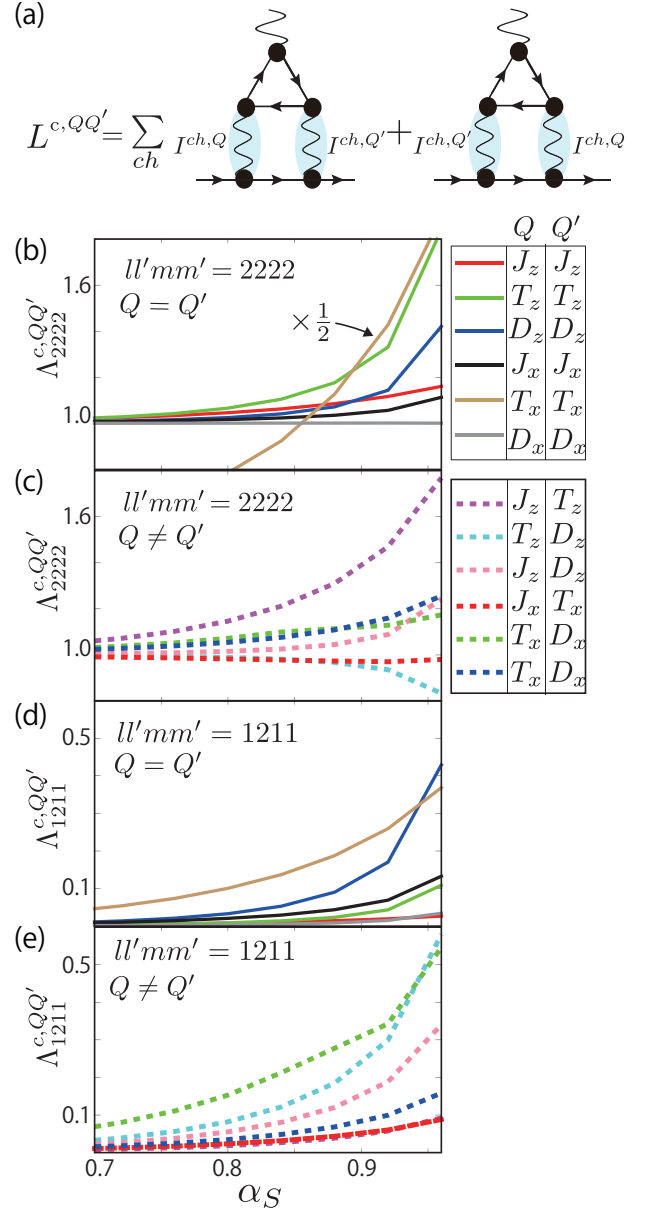}
\caption{(a) Multipole-decomposed $U$-VC given by $\Lambda^{c,QQ'}$.
$Q$ and $Q'$ are magnetic multipole operators.
Obtained $\hat{\Lambda}^{c,QQ'}_{2222}$ for (b) $Q=Q'$ and (c) $Q\neq Q'$, and
$\hat{\Lambda}^{c,QQ'}_{1211}$ for (d) $Q=Q'$ and
(e) $Q\neq Q'$. Many pairs of multipole fluctuations (Q,Q') contribute to the enhancement of $U$-VC.}
\label{fig:uvchitai2}
\end{figure}
\section{superconductivity}
\label{sec:super}
Now, we discuss about obtained superconducting states 
by solving the gap equation in Eq. (\ref{eqn:linear}).
The paring interaction is given by Eqs.(\ref{eqn:Vsc})-(\ref{eqn:Vchi}).
We solve the gap equation in the presence of both $u$ and $g$, by the following replacement,
\begin{eqnarray}
u\hat{U}^{0;c} \rightarrow u\hat{U}^{0;c}+2g\hat{W}
\label{eqn:replace}
\end{eqnarray}
in $\hat{I}^{c}(k-k')$ in the paring interaction (\ref{eqn:Vch}).
For finite $g$, $\hat{I}^{c}(\propto \hat{\chi}^{c})$ develops as large as $\hat{I}^{s}$ and $\hat{I}^{s\perp}$.
We put $g=0$ for $\hat{\Lambda}^{ch}$ approximately since the contribution from $\hat{\chi}^{c}$ remains small even for $g>0$ \cite{rina2}.

Figures \ref{fig:phase}(a)-(b) are obtained phase diagrams, 
which show the largest eigenvalue and its symmetry of the gap function. 
In the presence of $U$-VC, fully gapped $s$-wave state without any sign reversal 
emerges when $\alpha_{S}\lesssim 1$ and $\alpha_{C}\lesssim 1$
as shown in Fig.\ref{fig:phase}(a).
The region of $s$-wave phase gets wider as the 
magnetic fluctuations develop.
These results originate from the fact that
the charge-channel attractive interaction 
$-\frac{1}{2}\hat{V}^{c}$ are strongly enhanced by 
the charge-channel $U$-VC, which is enlarged due to the magnetic (odd-rank) 
multipole fluctuations when $\alpha_{S}\lesssim 1$.
In fact, $-\frac{1}{2}\hat{V}^{c}$ is expressed as
$-\frac{1}{2}\hat{V}^{c}\propto 
-\frac{1}{2}|\hat{\Lambda}^{c}|^{2}\{ (u-2g)^{2}\chi^{c}-(u-2g)\}$,
which takes large negative (=attractive) value when $\alpha_{C}\lesssim 1$
\cite{rina2}.
In addition, we find that quite small $g$
is enough for realizing the $s$-wave superconductivity.
For instance, $s$-wave state emerges even at $g=0.025$.
This is much smaller than Coulomb interaction $u=0.31$.

In contrast, the $s$-wave region in Fig.\ref{fig:phase}(a) is drastically reduced 
if we neglect $U$-VC ($\hat{\Lambda}^{ch}=\hat{1}$) as shown in Fig.\ref{fig:phase}(b).
In this case, $d_{x^{2}-y^{2}}$-wave state appears in wide parameter region.
Furthermore, the eigenvalue $\lambda$ for $d_{x^{2}-y^{2}}$-wave state in Fig.\ref{fig:phase}(b)
is much smaller than that for $s$-wave state in Fig.\ref{fig:phase}(a), so $T_{c}$ of
$d_{x^{2}-y^{2}}$-wave state should be very low if realized.
Therefore, we clearly confirmed that 
$U$-VC is important for realizing the $s$-wave superconductivity.
Obtained gap functions on Fermi surface for $s$- and $d_{x^{2}-y^{2}}$-wave states
are expressed in Fig.\ref{fig:phase}(c) and (d), respectively.
The obtained $s$-wave gap function is almost isotropic while the $d_{x^2-y^2}$-wave gap function has accidental nodes
in addition to the symmetry nodes.

 In conclusion, once the small electron-phonon interaction exist, 
fully gapped $s$-wave superconducting state
can appears in $f$-electron system near the magnetic QCP.
This counter-intuitive result is given by the large $U$-VC
 caused by multiple (higher-rank) multipole fluctuations.
We comment that the obtained large eigenvalues $\lambda$ in Fig.\ref{fig:phase} are overestimated
 since the self-energy effects (such as the mass-renormalization and the quasi particle damping) are dropped in the gap equation.

Finally, we show that multi-orbital nature is a necessary condition for realizing the
 $s$-wave superconductivity.
 In the present model, $f$-orbitals $| f_{1} \rangle$ and $| f_{2} \rangle$
have different itinerancy:
$| f_{1} \rangle$ is relatively itinerant and $| f_{2} \rangle$ is relatively localized.
We also introduce the CEF splitting $\Delta E$ between $| f_{1} \rangle$ and $| f_{2} \rangle$:
$E_{1}=E_{2}+\Delta E$
as shown in Fig.\ref{fig:gainen}(a).
In this model, the ratio between the $f$-orbital DoS at the Fermi level, $D^{f_{1}}(0)/D^{f_{2}}(0)$, is much larger than unity at
$\Delta E=0$, and the ratio decreases with $\Delta E$  as shown in Fig.\ref{fig:gainen}(b).
The ratio reaches unity at $\Delta E\simeq 0.12$.
In Fig.\ref{fig:gainen}(c) and (d), we show the obtained phase diagram at $\Delta E=0.06$
and  $\Delta E=0.12$, respectively.
The region of $s$-wave state at $\Delta E=0.12$ is much wider than that 
at $\Delta E=0.06$, which means that $s$-wave state is favored as $\Delta E$ increases.
As a result, the condition $D^{f_{1}}(0)\approx D^{f_{2}}(0)$ is significant for realizing 
the $s$-wave superconducting state.
In other words, the multi-orbital nature on Fermi surface is important for realizing
$s$-wave states.
Therefore, the $s$-wave state emerges in the presence of finite CEF splitting of $f$-levels
when the $s$-$f$ hybridization has strong orbital dependence.
This situation is expected to be realized in CeCu$_{2}$Si$_{2}$ at $P=0$\cite{LDADMFT_multiporbital}.
\begin{figure}[htb]
\includegraphics[width=.90\linewidth]{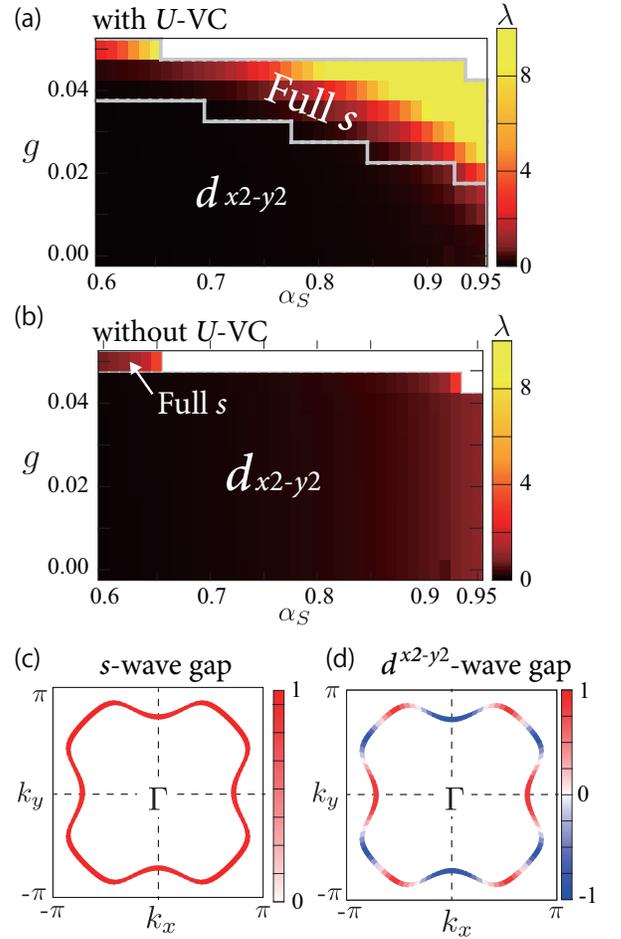}
\caption{(a) Phase diagram in the presence of $U$-VC.
The $s$-wave state emerges due to the significant contribution from $U$-VC. 
The white region corresponds to $\alpha_{C}>1$.
(b) Phase diagram in the absence of $U$-VC. Anisotropic $d_{x^2-y^2}$-wave state
appears in wide parameter region. The gap function on Fermi surface for (c) $s$-wave and (d) $d_{x^2-y^2}$-wave.}
\label
{fig:phase}
\end{figure}
\begin{figure}[htb]
\includegraphics[width=.96\linewidth]{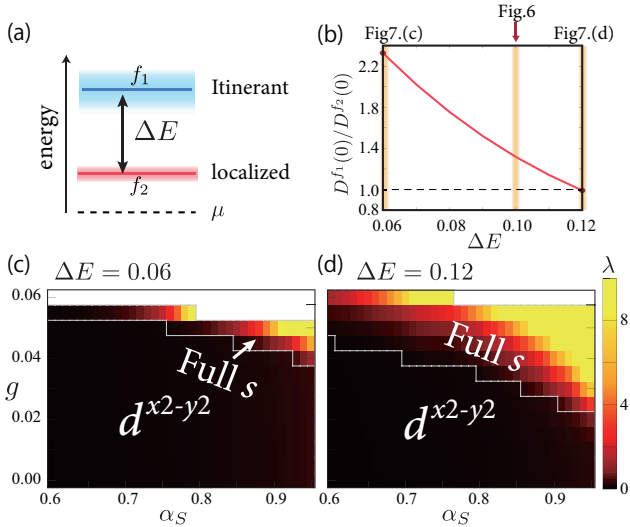}
\caption{(a) The energy level of the $f$-orbital states in the present model. 
(b) $\Delta E$ dependence of the ratio of the DoS
$D^{f_{1}}(0)/D^{f_{2}}(0)$.
The ratio goes to unity at $\Delta E \simeq 0.12$.
Obtained phase diagram at (c) $\Delta E =0.06$ and (d) $\Delta E =0.12$.}
\label{fig:gainen}
\end{figure}
\section{Summary}
In this paper,
we proposed a mechanism of $s$-wave superconductivity
in multi-orbital heavy fermion systems based on the recently developed 
beyond Migdal formalism.
In the present two-orbital PAM,
various odd-rank multipole fluctuations strongly develop simultaneously,
due to the combination of strong SOI and Coulomb interaction as shown in Fig.\ref{fig:chi}(b).
We verified that the result in Fig.\ref{fig:chi}(b) is qualitatively same as those for $\Delta E=0\sim 0.2$. 
These developed fluctuations give significant $U$-VCs,
by which the model Coulomb interaction is strongly modified.
Especially, the coupling constant
between electron and charged-boson (=$u\hat{U}^{0,c}+2g\hat{W}$)
is prominently magnified by the $U$-VC as shown in Figs.\ref{fig:uvchitai1} and \ref{fig:uvchitai2}.
For this reason, even-rank multipole fluctuations
give large attractive interaction
when the system is close to the magnetic QCP.
We revealed that
$s$-wave superconductivity is strongly enhanced 
near the magnetic criticality in multi-orbital heavy fermion systems,
once moderate phonon-induced multipole fluctuations exist as shown in Fig.\ref{fig:phase}.
Note that the depairing effect of Coulomb interaction 
is reduced by the multiorbital screening effect \cite{rina2}.
The present mechanism may be responsible for the
fully gapped $s$-wave superconducting state
realized in CeCu$_2$Si$_2$.

The main results of the present study on the two-orbital
periodic Anderson model with strong SOI are summarized as follows:
near the magnetic QCP, we find that
(i) several (higher-rank) multipole fluctuations strongly develop simultaneously, 
whereas rank-1 orbital-diagonal spin susceptibility solely develops in 3$d$-electron systems.
(ii)  Multiple multipole fluctuations give large $U$-VC cooperatively, 
leading to the violation of Migdal theorem.
(iii)  Thanks to $U$-VC, electric-multipole fluctuation mediated $s$-wave 
superconductivity is realized
when $D^{f1}(0) \approx D^{f2}(0)$, which is a necessary
condition for realizing moderate quadrupole or hexadecapole fluctuations.

In this study, we introduced a phenomenological
interaction term in Eq. (\ref{eqn:ph}) in order to realize the moderate
$A^{+}_{1}$-channel multipole fluctuations.
This term can originate from moderate electron-phonon interaction, 
as we discussed in the main text.
In fact, 
strong coupling between $f$-electrons and phonons
via the $s$-$f$ hybridization and $f$-orbital level
is expected in heavy fermion systems,
as discussed in Refs.\cite{Razaf,Ohkawa,Nagaoka},
For example, large Gruneisen parameter in heavy fermion systems
($\eta\equiv -d{\rm log}T_K/d{\rm log}\Omega \sim 30-80$) 
indicates the significance of electron-phonon interaction \cite{Razaf}.
The phonon-mediated $s$-wave superconductivity 
in heavy fermion systems discussed in 
Refs.\cite{Razaf,Ohkawa,Nagaoka}
becomes a realistic scenario by considering the significant role of
$U$-VC revealed in the present study.
Another promising microscopic origin of Eq. (\ref{eqn:ph}) is
the AL-type VCs for the susceptibility.
In fact, in 3$d$-electron systems without SOI,
the AL-type VCs causes large orbital fluctuations
\cite{Onari-SCVC}.
Recently, the present authors found that the
AL-type VCs give large even-rank multipole fluctuations
in heavy-fermion systems with strong SOI,
which we will discuss in future publication
\cite{Tazai-future}.

There are many important future issues.
For example, it is interesting to apply the present theory
to more realistic three-dimensional model for CeCu$_2$Si$_2$.
Also, study of self-energy correction,
which gives strong mass-enhancement
whereas neglected in the present study,
is an important future issue.
In addition, the present theory may be applicable for spin-triplet
superconductor UPt$_3$.
In fact, we analyzed the multiorbital Hubbard model
for Sr$_2$RuO$_4$, and found that the triplet state
is realized under the coexistence of
spin and orbital fluctuations
\cite{rina1,Tsuchiizu}.

\acknowledgements
We are grateful to Y. Matsuda, T. Shibauchi, Y. Kasahara, S. Kittaka,
H. Ikeda, S. Onari and Y. Yamakawa for useful discussions.
This study has been supported by KAKENHI Grant Numbers 
(No.18J12852, No. 18H01175) from the Japan Society for the Promotion
of Science (JSPS).

\section*{Appendix A: $s$-$f$ Hybridization}
\label{appendix1}
Here, we derive the expression of $s$-$f$ hybridization given in Eq. (\ref{eqn:hopping}).
In the LS basis, the wave function of
$f$-electrons  in Eq. (\ref{eqn:wavefunc}) are given by
\begin{eqnarray}
|f_{1}\Uparrow \rangle
&=&a \left\{ \sqrt{\frac{6}{7}}| -3,\uparrow \rangle - \sqrt{\frac{1}{7}}
 |-2,\downarrow \rangle \right\} \nonumber \\
&&+b \left\{ \sqrt{\frac{2}{7}}| 1,\uparrow \rangle - \sqrt{\frac{5}{7}} |2,\downarrow \rangle \right\},\label{eqn:wavefunc2-1} \\ 
|f_{1}\Downarrow \rangle 
&=&a \left\{ \sqrt{\frac{1}{7}}| 2,\uparrow \rangle - \sqrt{\frac{6}{7}}
 |3,\downarrow \rangle \right\} \nonumber \\
&&+b \left\{ \sqrt{\frac{5}{7}}| -2,\uparrow \rangle - \sqrt{\frac{2}{7}} |-1,\downarrow \rangle \right\},\label{eqn:wavefunc2-2} \\
|f_{2}\Uparrow \rangle
&=&-a \left\{ \sqrt{\frac{2}{7}}| 1,\uparrow \rangle - \sqrt{\frac{5}{7}} |2,\downarrow \rangle \right\} \nonumber \\
&&+b \left\{ \sqrt{\frac{6}{7}}| -3,\uparrow \rangle - \sqrt{\frac{1}{7}}
 |-2,\downarrow \rangle \right\},\label{eqn:wavefunc2-3} \\ 
|f_{2}\Downarrow \rangle
&=&-a \left\{ \sqrt{\frac{5}{7}}| -2,\uparrow \rangle - \sqrt{\frac{2}{7}} |-1,\downarrow \rangle \right\} \nonumber \\
&&+b \left\{ \sqrt{\frac{1}{7}}| 2,\uparrow \rangle - \sqrt{\frac{6}{7}}
 |3,\downarrow \rangle \right\},
\label{eqn:wavefunc2-4}
\end{eqnarray}
where $\uparrow(\downarrow)$ is the real spin.
Note that the wave functions for $L_{z}=\pm 2$ are proportinal to z as follows,
\begin{eqnarray}
\langle \vec{r} \,\, |\pm 2,\sigma \rangle \propto z. \nonumber
\label{eqn:lz2}
\end{eqnarray}
Now, we consider the hybridization between $f$-electrons in
Eq. (\ref{eqn:wavefunc2-1})-(\ref{eqn:wavefunc2-4}) and $s$-electron.
In 2D system, the hybridization between $s$-orbital at $\vec{R}_{i}$ site and 
$|\pm 2,\sigma \rangle$ at $\vec{R}_{j}$ site goes to zero, that is, 
$\langle s, \sigma, \vec{R}_{i} |\pm 2, \sigma, \vec{R}_{j} \rangle =0$.
Then, we obtain
\begin{eqnarray}
\langle s \uparrow|f_{1}\Uparrow \rangle
&=&  \sqrt{\frac{6}{7}}\langle s \uparrow | -3,\uparrow \rangle,\\
\langle s \downarrow|f_{1}\Downarrow \rangle 
&=&- \sqrt{\frac{6}{7}}\langle s \downarrow | 3,\downarrow \rangle, \\
\langle s \uparrow|f_{2}\Uparrow \rangle
&=&- \sqrt{\frac{2}{7}}\langle s \uparrow | 1,\uparrow \rangle, \\
\langle s \downarrow|f_{2}\Downarrow \rangle
&=& \sqrt{\frac{2}{7}}\langle s \downarrow | -1,\downarrow \rangle,
\label{eqn:wavefunc3}
\end{eqnarray}
where $a=1$ and $b=0$. Therefore, we obtain the relation $\langle s \uparrow|f_{l}\Downarrow \rangle = 
\langle s \downarrow|f_{l}\Uparrow \rangle =0$.
As a results, we confirm that 
the pseudo-spin is conserved in the $s$-$f$ hybridization in the present two-orbital model.
Therefore, we can use the pseudo-spin channel $(s,s\!\perp,c)$,
in the present study, which is a great merit for performing detailed analysis.
\section*{Appendix B: Coulomb interaction}
\label{appendix2}
Here, we explain about the Coulomb interaction in TABLE I in more detail. 
The Coulomb interaction is obtained by the following steps.
Firsts, we calculate the $L_{z}$-basis-Coulomb interaction $\bar{U}_{l_{z},l'_{z},l''_{z},l'''_{z}}$ 
by using Eq. (\ref{eqn:lzcoulomb}).
The obtained Coulomb interaction is written by using the Slater integral parameters 
$(F_{0},F_{2},F_{4},F_{6})$.
Note that $\bar{U}_{l_{z},l'_{z},l''_{z},l'''_{z}}=0$ for
$l_{z}+l'''_{z}\neq l'_{z}+l''_{z}$.
Next, we transfer it from the $L_{z}$-basis into the $L=(l,\sigma)$ basis, which is 
given by the unitary transformation from the right-hand to the left-hand parts in Eqs.(\ref{eqn:wavefunc2-1})-(\ref{eqn:wavefunc2-4}).
The obtained Coulomb interaction satisfies the axial rotational symmetry expressed as Eq. (\ref{eqn:U0sc})
after antisymmetrization.
In the case of $a=1$ and $b=0$, the obtained Coulomb interaction is written by using the
$U^{1},U^{2},J,J^{\perp},J', J^{x1},$ and $J^{x2}$.
The definition of each element is given in Fig.\ref{fig:coulomb}(a), and 
the obtained values are shown  in Fig.\ref{fig:coulomb}(b).
Although the other elements not listed in Fig.\ref{fig:coulomb} (e.g., $U^{0;ch}_{11;12}$)
are zero at $a=1$, they become finite for $a\lesssim 1$. 
Note that, in 3$d$-electron systems without SOI, the relations $J=J^{\perp}$ and $J^{x1}=J^{x2}=0$ are
satisfied.

Finally, TABLE I is obtained by introducing the anti-symmetrization of the Coulomb interaction. The TABLE I becomes equal to the table of Coulomb interaction 
in 3d-electron systems \cite{rina2} if we put $J=J^{\perp}$ and $J^{x1}=J^{x2}=0$ in TABLE I. 
We stress that the pseudo-spin is conserved even for $a\neq1$.
\begin{figure}[htb]
\includegraphics[width=0.94\linewidth]{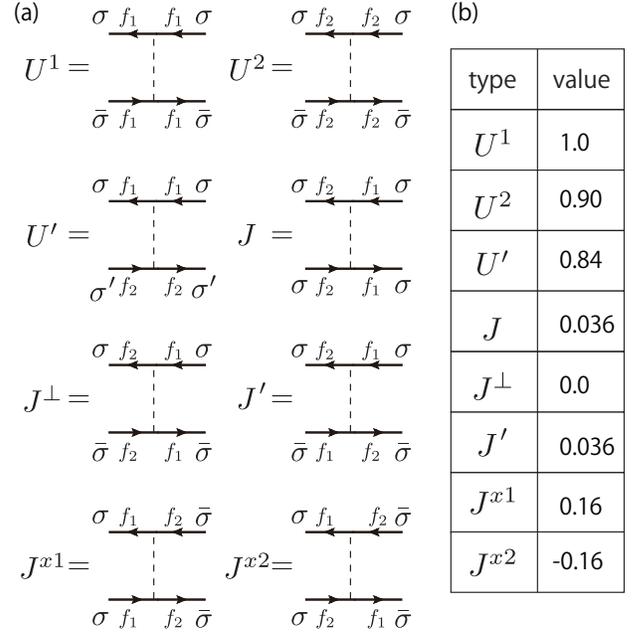}
\caption{(a) Definition of multi-orbital Coulomb interaction in the pseudo-spin representation; $U^{1},U^{2},J,J^{\perp},J', J^{x1},$ and $J^{x2}$. 
(b) Obtained value for the Coulomb interaction when $a=1$ and $(F_{0},F_{2},F_{4},F_{6})=(5.3,9.09,6.927,4.756)$ in unit eV. These values are normalized as $U^{1}=1.0$.
 (Before the normalization, $U^{1}=6.1$ eV.)}
\label{fig:coulomb}
\end{figure}
\section*{Appendix C: Multipole operator}
\label{appendix3}
Here, we explain about the multipole operators in TABLE II. 
We numerically obtain each operators 
by using $4\times4$ tensor $J^{(k)}_{q}$ in Eq. (\ref{eqn:J3}).
As a result, electric (even-rank) multipole operators in $D_{4h}$ symmetry are expressed as
\begin{eqnarray}A_{1}^{+}&&
\begin{cases}
\hat{1}&=\hat{\sigma}^{0}\hat{\tau}^{0}  \nonumber \\
\hat{O}_{20}&=\hat{\sigma}^{0} \left(2.00\hat{\tau}^{0}+3.00\hat{\tau}^{z} \right)  \nonumber \\
\hat{H}_{0}&=\hat{\sigma}^{0} \left(-5.73\hat{\tau}^{0}+11.5\hat{\tau}^{z}-12.8\hat{\tau}^{x} \right)
\end{cases} \nonumber \\ A_{2}^{+}&&
\begin{cases}
\hat{H}^{z}&=-19.8 \hat{\sigma}^{z} \hat{\tau}^{y}  
\end{cases} \nonumber \\ E^{+}&&
\begin{cases}
\hat{O}_{yz}&=-3.87 \hat{\sigma}^{x} \hat{\tau}^{y}  \nonumber \\
\hat{O}_{zx}&=+3.87 \hat{\sigma}^{y} \hat{\tau}^{y}  \nonumber \\
\end{cases} \label{eqn:eleO}\\
\end{eqnarray}
Magnetic (odd-rank) multipole operators are given by
\begin{eqnarray} A_{1}^{-}&&
\begin{cases}
\hat{D}_{4}&=+29.8i \hat{\sigma}^{0} \hat{\tau}^{y}  \nonumber \\
\end{cases} \nonumber \\ A_{2}^{-}&&
\begin{cases}
\hat{J}^{z}&= \hat{\sigma}^{z} \left(0.50\hat{\tau}^{0}+2.00\hat{\tau}^{z} \right)   \nonumber \\
\hat{T}^{z}&= \hat{\sigma}^{z} \left(9.00\hat{\tau}^{0}-1.50\hat{\tau}^{z} \right)   \nonumber \\
\hat{D}^{z}&= -29.8 \hat{\sigma}^{z} \hat{\tau}^{x} 
\end{cases}  \\
E^{-}&&
\begin{cases}
\hat{J}^{x}&= -1.12 \hat{\sigma}^{x} \hat{\tau}^{x}   \nonumber \\
\hat{J}^{y}&= -1.12 \hat{\sigma}^{y} \hat{\tau}^{x}   \nonumber \\
\hat{T}^{x}&=  \hat{\sigma}^{x}  \left(3.75\hat{\tau}^{0}-3.75\hat{\tau}^{z} +5.03\hat{\tau}^{x}\right)   \nonumber \\
\hat{T}^{y}&=  \hat{\sigma}^{y}  \left(3.75\hat{\tau}^{0}-3.75\hat{\tau}^{z} +5.03\hat{\tau}^{x}\right)   \nonumber \\
\hat{D}^{x}&=  \hat{\sigma}^{x}  \left( 23.0\hat{\tau}^{0}-6.56\hat{\tau}^{z}-3.14\hat{\tau}^{x}\right)   \nonumber \\
\hat{D}^{y}&=  \hat{\sigma}^{y}  \left( 23.0\hat{\tau}^{0}-6.56\hat{\tau}^{z}-3.14\hat{\tau}^{x}\right)  
\end{cases} \label{eqn:magneO} \\
\end{eqnarray}
where $\hat{\sigma}^{\mu}$ and $\hat{\tau}^{\mu}$($\mu=x,y,z$) are Pauli matrices for
 the pseudo-spin and orbital basis, respectively.
$\hat{\sigma}^{0}$ and $\hat{\tau}^{0}$ are identity matrices.
We express the obtained 16 matrix expressions in Eq. (\ref{eqn:eleO}) and (\ref{eqn:eleO})
as $O^{Q}(Q=(\Gamma,\phi))$. Note that $\sum_{LL'} O^{(\Gamma,\phi)}_{LL'}
O^{(\Gamma',\phi') *}_{LL'}=0$ for $\Gamma \neq \Gamma'$, whereas 
$\sum_{LL'} O^{(\Gamma,\phi)}_{LL'} O^{(\Gamma,\phi')*}_{LL'}\neq 0$
$A^{+}_{2}$($E^{+}$) electric multipole operators
belong to pseudo-spin $s\,(s\!\!\perp)$ channel since it is proportional to 
$\hat{\sigma}^{z} (\hat{\sigma}^{x},\hat{\sigma}^{y})$.
Also, $A^{-}_{1}$ magnetic multipole operators belong to the charge channel since it is proportional to $\hat{\sigma}^{0}$. The 
In summary, some electric (magnetic) multipole
operators belong to pseudo-spin (charge) channels as summarized in TABLE II.
The relation between multipole and pseudo-spin (charge) channels
We have to take care of this fact in analysis.

\section*{Appendix D: Effects of $f$-$f$ hopping}
\label{appendix5}
In this section, we discuss about the effects of $f$-$f$ hopping.
In the main text, we neglected $f$-$f$ hopping, and therefore the $f_{l}$-orbital weight is quite isotropic on Fermi surface as shown in Fig.\ref{fig:band}(e).
However, this orbital-isotropy can be broken if we introduce finite $f$-$f$ hopping.
Now, we introduce the orbital-dependent $f$-$f$ hopping. 
In this case, $f$-electron energy $E_{l}$ have $\k$-dependence.
As a result, the $f_{l}$-orbital weight comes to have $\theta$-dependence 
on the Fermi surface. The $f$-$f$ hopping is expressed as
\begin{eqnarray} \hat{H}_{ff}=\sum_{\k l\sigma}E_{\k,l}f^{\dagger}_{\k l\sigma}
f_{\k l\sigma}.\end{eqnarray}
Here, we set $E_{\k,1}\equiv E_{1}+\delta E_{\k}$ and $E_{\k,2}\equiv E_{2}-\delta E_{\k}$,
where the $\k$-dependence of $\delta E_{\k}$ is shown in Fig.\ref{fig:appendix}(a).
Technically, to realize the $\delta E_{\k}$, we introduce the intra-orbital $f$-$f$ hopping 
up to fifth nearest neighbor hopping integrals according to Ref.\cite{Yamakawa-FeSe2}. 
In Fig.\ref{fig:appendix}(b), we show the obtained $f_{l}$-orbital weight along 
$\theta$-axis on Fermi surface.
It shows strong $\theta$-dependence irrespective of the fact that $|\delta E_{k}|(\sim 0.2)$ is much smaller than $|t_{sf}|(=0.7)$.

One may suspect that higher rank multipole susceptibilities may be
suppressed when the $f$-orbital weight is $\theta$-dependent, since 
the orbital off-diagonal components of $\chi^{s}_{ll'mm'}$ 
may be suppressed .
To answer this question, we perform the RPA analysis.
Figure \ref{fig:appendix}(c) shows the obtained magnetic multipole susceptibilities.
We find that multiple higher-rank magnetic multipole 
susceptibilities develop, which is quite similar to our result without $f$-$f$ hopping in Fig.\ref{fig:chi}(b). 
This unexpected results originate from the fact that many body effects away from Fermi 
energy also contribute to the multipole susceptibility.
This result strongly indicates that $U$-VC is still important 
even in the presence of small $f$-$f$ hopping. 
We study this issue in more detail in the future publication \cite{Tazai-future}
\begin{figure}[htb]
\includegraphics[width=0.96\linewidth]{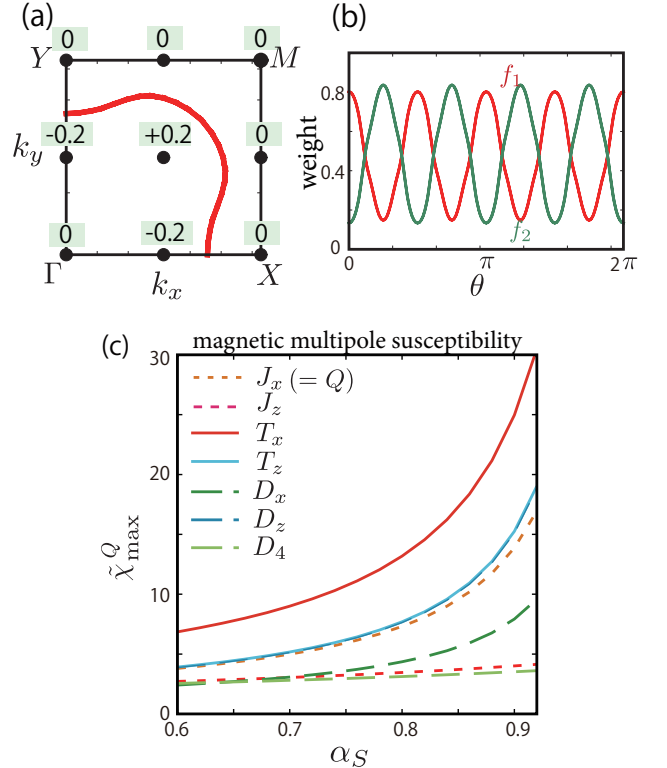}
\caption{(a) The Fermi surface with $f$-$f$ hopping.
Each number at $\k$ shows intra-orbital energy shift $\delta E_{\k}$.
 (b) Obtained $\theta$-dependence of the 
$f_{l}$-orbital weight on Fermi surface. The red (green) line corresponds to $f_{1}$($f_{2}$)-orbital. 
(c) $\a_{S}$ dependence of magnetic multipole susceptibilities, which are almost equal to those in
Fig.\ref{fig:chi}(b).}
\label{fig:appendix}
\end{figure}
\section*{Appendix E: Multipole expansion}
\label{appendix4}
In this section, we explain about the derivation of the coefficient $a^{\Gamma, \phi, \phi'}$ in Eq. (\ref{eqn:shaei2}).
First, we solve the characteristic equation for the $f$-electron susceptibility in the $L=(l,\sigma)$ basis,
\begin{eqnarray}
\sum_{MM'} \chi_{LL'MM'}(q) v_{MM'}^{i}(q) = \lambda^{i}(q) v_{LL'}^{i}(q),
\end{eqnarray}
where  $\lambda^{i}(q)$ is i-th real eigen value ($i=1\sim 16$). 
$v^{i}_{LL'}(q)$ is a $16$-dimensional eigen vector.
In the present model, $\chi^{(\Gamma,\phi),(\Gamma,\phi')}=0$ for $\Gamma \neq \Gamma'$. Thus, for each $i$, $v^{i}_{LL'}(q)$ is classified into the corresponding IR ($\Gamma$).
If we normalize $\vec{v}^{i}$ as $\sum_{LL'}v^{i}_{LL'}(v^{j}_{LL'})^{*}=\delta_{i,j}$,
the $f$-electron susceptibility is expressed as  
\begin{eqnarray}
\chi_{LL'MM'}(q) =\sum_{i}v_{LL'}^{i}(q)\lambda^{i}(q) v_{MM'}^{i}(q)^{*}
\label{eqn:bdef2}
\end{eqnarray}
Then, we expand  $\vec{v}^{i}(q)$ for $i \in \Gamma$ on the basis of the multipole matrices
$\hat{O}^{(\Gamma,\phi)}$ for $\phi=1\sim N_{\Gamma}$ listed in Eqs.(\ref{eqn:eleO}) and (\ref{eqn:magneO}) as follows:
\begin{eqnarray}
v_{LL'}^{i}(q)=\sum_{\phi=1}^{N_{\Gamma}}
b^{i,\phi}(q)
O_{LL'}^{(\Gamma,\phi)},
\label{eqn:bdef}
\end{eqnarray}
where the coefficient $b^{i,\phi}(q)$ is uniquely determined.
Note that the basis $\{ \vec{O}^{(\Gamma,\phi)} \}$ is complete but not orthogonal within the
same $\Gamma$.
By inserting Eq. (\ref{eqn:bdef}) into Eq. (\ref{eqn:bdef2}), we obtain
\begin{eqnarray}
\chi_{LL'MM'}(q) =
\sum_{\Gamma,\phi,\phi'}
a^{\Gamma,\phi,\phi'}(q)
O_{LL'}^{(\Gamma,\phi)}
O_{MM'}^{(\Gamma,\phi')*},
\end{eqnarray}where
\begin{eqnarray}
a^{\Gamma,\phi,\phi'}(q)=\sum_{i\in \Gamma}
b^{i,\phi}(q)
\lambda^{i}(q)
b^{i,\phi'}(q)^{*}.
\end{eqnarray}
As a result, the decomposition of $\chi_{LL'MM'}(q)$ in Eq. (\ref{eqn:shaei2}) is obtained.
In the same way, the paring interaction $I$ in Eq. (\ref{eqn:shaei3}) can be decomposed.
Using $a^{\Gamma,\phi,\phi'}$, the multipole susceptibility $\chi^{(\Gamma,\phi),(\Gamma,\phi')}$ defined in Eq. (\ref{eq:suscep}) is expressed as
\begin{eqnarray}
\chi^{(\Gamma,\phi),(\Gamma,\phi')}=\sum_{\phi'' \phi'''} a^{\Gamma, \phi'', \phi'''}
(T^{\Gamma}_{\phi, \phi''})^{*} T^{\Gamma}_{\phi', \phi'''},
\end{eqnarray} where
\begin{eqnarray}
T^{\Gamma}_{\phi, \phi'}=\sum_{MM'} O^{(\Gamma, \phi)}_{MM'}
(O^{(\Gamma, \phi')}_{MM'})^{*}.
\end{eqnarray}

\end{document}